\documentclass[a4paper,11pt]{article}

\pdfoutput=1

\usepackage{jheppub}
\usepackage{graphicx}
\usepackage{amsmath}
\usepackage{amssymb}
\usepackage{mathrsfs}
\usepackage{bm}
\usepackage{xspace}
\usepackage{mdwlist}
\usepackage[dvipsnames]{xcolor}
\usepackage[normalem]{ulem}
\usepackage{slashed}
\usepackage{hyperref}

\newcommand{\eq}[1]{eq.~\eqref{eq:#1}}
\newcommand{\eqs}[2]{eqs.~\eqref{eq:#1} and \eqref{eq:#2}}
\renewcommand{\sec}[1]{sec.~\ref{sec:#1}}

\newcommand{\fig}[1]{fig.~\ref{fig:#1}}

\newcommand{\lb}{\Big{\lbrack}}
\newcommand{\rb}{\Big{\rbrack}}
\newcommand{\lp}{\Big{(}}
\newcommand{\rp}{\Big{)}}

\newcommand{\nn}{\nonumber}
\newcommand{\nl}{\nonumber \\}

\renewcommand{\pmb}[1]{\boldsymbol{#1}}
\newcommand{\bmat}[1]{\boldsymbol{#1}}

\newcommand{\ord}[1]{{\mathcal O}(#1)}

\newcommand{\df}{\mathrm{d}}  
\newcommand{\img}{\mathrm{i}}

\newcommand{\al}{\alpha}

\newcommand{\ga}{\gamma}
\newcommand{\Ga}{\Gamma}
\newcommand{\de}{\delta}
\newcommand{\De}{\Delta}

\newcommand{\la}{\lambda}
\newcommand{\si}{\sigma}
\renewcommand{\t}{\tau}

\newcommand{\CS}{\mathcal{C}}

\newcommand{\cusp}{\text{cusp}}

\newcommand{\bn}{\bar{n}}

\allowdisplaybreaks[2]

\title{Joint thrust and TMD resummation in electron-positron and electron-proton collisions}

\author[a]{Yiannis Makris,}
\author[b]{Felix Ringer,}
\author[c,d]{Wouter J. Waalewijn}

\affiliation[a]{INFN Sezione di Pavia, via Bassi 6, I-27100 Pavia, Italy}
\affiliation[b]{Nuclear Science Division, Lawrence Berkeley National Laboratory, Berkeley, CA 94720, USA}
\affiliation[c]{Institute for Theoretical Physics Amsterdam and Delta Institute for Theoretical Physics, University of Amsterdam, Science Park 904, 1098 XH Amsterdam, The Netherlands}
\affiliation[d]{Nikhef, Theory Group, Science Park 105, 1098 XG, Amsterdam, The Netherlands}
     
\emailAdd{yiannis.makris@pv.infn.it}
\emailAdd{fmringer@lbl.gov}
\emailAdd{w.j.waalewijn@uva.nl}

\abstract{We present the framework for obtaining precise predictions for the transverse momentum of hadrons with respect to the thrust axis in $e^+e^-$ collisions. This will enable a precise extraction of transverse momentum dependent (TMD) fragmentation functions from a recent measurement by the Belle Collaboration.  Our analysis takes into account, for the first time, the nontrivial interplay between the hadron transverse momentum and the cut on the thrust event shape. To this end, we identify three different kinematic regions, derive the corresponding factorization theorems within Soft Collinear Effective Theory, and present all ingredients needed for the joint resummation of the transverse momentum and thrust spectrum at NNLL accuracy. One kinematic region can give rise to non-global logarithms (NGLs), and we describe how to include the leading NGLs. We also discuss alternative measurements in $e^+e^-$ collisions that can be used to access the TMD fragmentation function. Finally, by using crossing symmetry, we obtain a new way to constrain TMD parton distributions, by measuring the displacement of the thrust axis in $ep$ collisions.}

\preprint{NIKHEF 20-032}

\begin{document}
\maketitle

\section{Introduction}
\label{sec:intro}

Quarks and gluons produced in short-distance scattering processes fragment into hadrons at large distances, due to the confining nature of QCD. The fraction of the longitudinal momentum of the initial parton carried by these hadrons is described by fragmentation functions (FFs)~\cite{Georgi:1977mg,Ellis:1978ty,Curci:1980uw,Collins:1981uw}. FFs are independent of the scattering process, and enter in the factorization theorem of the corresponding cross section. The (typically small) difference between the direction of the hadron and initial parton it fragments from, can be taken into account by transverse momentum dependent (TMD) FFs~\cite{Collins:1992kk,Mulders:1995dh,Boer:1997nt,Boer:1997qn,Ji:2004wu,Collins:2011zzd}. See refs.~\cite{Metz:2016swz,Anselmino:2020vlp} for reviews of (TMD) FFs. TMD FFs are the final-state counter parts of TMD parton distribution functions (PDFs), which describe the intrinsic transverse momentum of partons inside nucleons. 

 Various determinations of TMD PDFs and FFs from experimental data have been performed: global extractions of unpolarized TMD PDFs and FFs~\cite{Sun:2013hua,Bacchetta:2017gcc,Scimemi:2019cmh}, extractions of TMD PDFs~\cite{Bertone:2019nxa,Bacchetta:2019sam}, and polarized TMD PDFs and FFs~\cite{Echevarria:2014xaa,Anselmino:2015sxa,Kang:2015msa,Callos:2020qtu,Cammarota:2020qcw}. One  process that is used for these extractions is hadron-pair production in $e^+e^-$~annihilation, for which the factorization theorem of the cross section differential in the relative transverse momentum of the two hadrons has the following schematic form, $\sigma \sim H D D S$. This involves a TMD FF for each of the identified hadrons, denoted by $D$, while the hard function $H$ describes the hard scattering and the TMD soft function $S$ encodes the effect of soft radiation.\footnote{These TMD FFs are referred to as ``unsubtracted", as one typically absorbs the square root of the TMD soft function into each of the TMD FFs.} Another process is SIDIS, where a single hadron is identified in the final state. The cross section differential in the hadron transverse momentum factorizes as $\sigma \sim H F D S$, where $F$ is the TMD PDF for the nucleon the electron scatters off. To disentangle the contribution from TMD FFs and TMD PDFs in SIDIS, a reliable understanding of TMD FFs from $e^+e^-$ data is of tremendous value. Consequently, this is critical to the success of the science program of the Electron-Ion Collider (EIC), where the extraction of TMD PDFs is one of the main goals~\cite{Accardi:2012qut}. Alternatively, measurements involving jets instead of final-state hadrons have been proposed in recent years to extract TMD PDFs in SIDIS~\cite{Gutierrez-Reyes:2018qez,Liu:2018trl,Gutierrez-Reyes:2019vbx,Gutierrez-Reyes:2019msa,Arratia:2020nxw}.

\begin{figure}[t!]
  \centerline{\includegraphics[width = \textwidth]{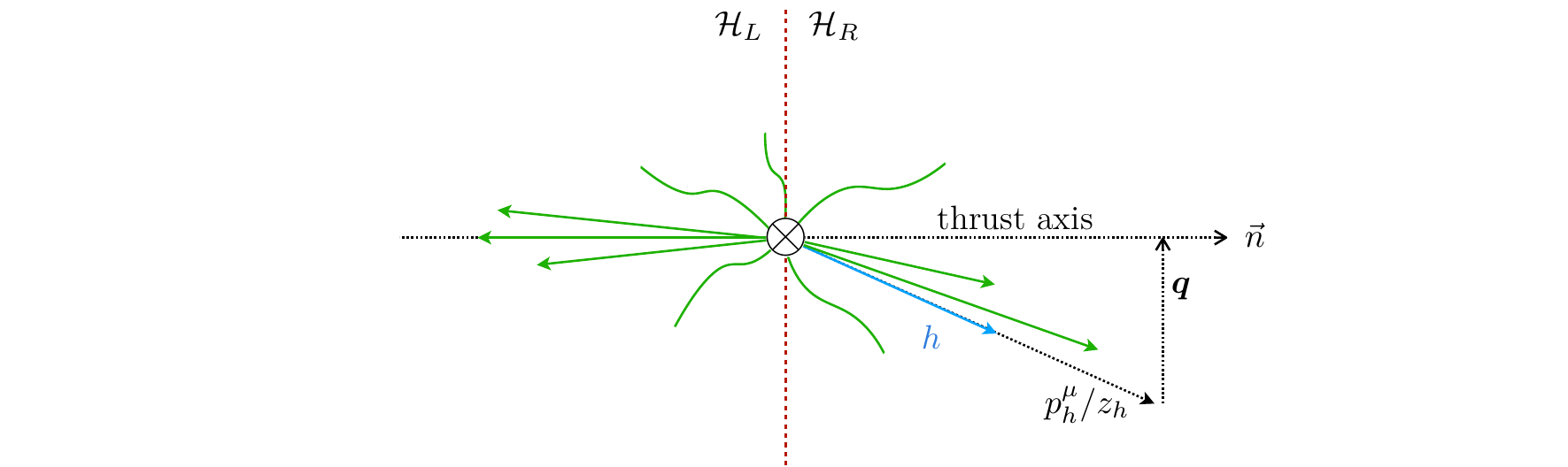}}
  \caption{Illustration of the measurement of the transverse momentum $\pmb{q} = \pmb{p}_h/z_h$ of a hadron (blue) relative to the thrust axis (black dashed) in an $e^+e^-$ collision denoted by $\otimes$. The plane perpendicular to the thrust axis (red dashed) divides the space into two hemispheres: $\mathcal{H}_L$ and $\mathcal{H}_R$.~\label{fig:measurement}}
\end{figure}

Recently, the Belle Collaboration measured the transverse momentum of hadrons in electron-positron annihilation relative to the thrust axis~\cite{Seidl:2019jei}, illustrated in \fig{measurement}. The thrust of an $e^+e^-$ event is defined as~\cite{Farhi:1977sg}
\begin{equation} \label{eq:thrust}
    1-\tau=T= \max _{\hat{t}} \frac{\sum_{i}\left|\hat{t} \cdot \vec{p}_{i}\right|}{\sum_{i}\left|\vec{p}_{i}\right|}\,,
\end{equation}
where the sum runs over all particles $i$ in the final state with 3-momentum $\vec p_i$. The unit vector vector $\hat t$ that maximizes $T$, defines the direction of the thrust axis, and naturally divides the final state into two hemispheres depending on the sign of $\hat{t} \cdot \vec{p}_{i}$ (see \fig{measurement}). Most events have $\tau \ll 1$, and in this case the final state consists of two highly-collimated back-to-back jets. The maximum value of $\tau=1/2$ corresponds to spherically symmetric events. The cross section reported in ref.~\cite{Seidl:2019jei} is differential in 
\begin{equation} \label{eq:meas}
    z_h=2E_h/Q\,,\quad p_{hT} \equiv |\pmb{p}_h| \,,\quad \tau \,,
\end{equation}
where $E_h$ is the hadron energy, $Q$ is the center-of-mass energy,  and $p_{hT}$ is the transverse momentum of the identified hadron.  The results from Belle are of great interest in order to study TMD evolution effects, 
which receive a nonperturbative contribution through the \emph{universal} rapidity anomalous dimension (Collins-Soper kernel),\footnote{Currently the nonperturbative contribution to the rapidity anomalous dimension is constrained by fits to experimental data~\cite{Landry:1999an,Landry:2002ix,Konychev:2005iy,DAlesio:2014mrz,Bacchetta:2017gcc,Scimemi:2017etj}, and methods for extracting it from lattice QCD are being explored~\cite{Ebert:2018gzl,Shanahan:2020zxr}} and test universality of TMDs by comparing with other processes, such as SIDIS, Drell-Yan or even jet substructure measurements. Furthermore, it will be interesting to test perturbative QCD at the lower energy of the Belle experiment, where nonperturbative effects are larger. Their high statistics allow for more precise and more differential measurements, such as the one we explore here.

The first factorization theorem for the cross section differential in the observables in \eq{meas} assumed $p_{hT}/z_h \sim \sqrt{\tau} Q\ll Q$~\cite{Jain:2011iu}, which corresponds to one of the kinematic regions we consider. Phenomenological studies related to the Belle measurement were performed in refs.~\cite{Boglione:2017jlh,Soleymaninia:2019jqo}, and the breaking of universality of nonperturbative soft effects between dihadron and single hadron production in $e^+e^-$ was discussed in ref.~\cite{Boglione:2020cwn}. Recently, ref.~\cite{Kang:2020yqw} proposed integrating out the thrust variable $\tau$,
and carried out the resummation of transverse momentum $p_{hT}$ and threshold logarithms $1-z_h$. This cross section contains non-global logarithms (NGLs)~\cite{Dasgupta:2001sh} because the transverse momentum measurement only constrains radiation in one hemisphere. The perturbative accuracy of observables that require the resummation of NGLs is currently limited to next-to-leading logarithmic (NLL) accuracy. We will also encounter NGLs, but in our analysis that is differential in thrust, they only appear in a restricted and clearly identified region of phase space. 

In this paper we derive the theoretical framework that is needed for the accurate extraction of TMD FFs from the cross section measured by the Belle Collaboration, crucially taking correlations between (the cut on) thrust and the transverse momentum measurement into account. 
These correlations enable us to identify the correct matrix elements and proper factorization of the cross section.
Specifically, we find three different kinematic regions and derive the factorization theorems for the corresponding cross section using Soft Collinear Effective Theory  (SCET)~\cite{Bauer:2000ew,Bauer:2000yr,Bauer:2001yt, Bauer:2002nz, Beneke:2002ph}. An extension of SCET, called SCET$_+$~\cite{Bauer:2011uc,Procura:2014cba,Larkoski:2015zka}, is required to jointly resum the large logarithmic corrections in \emph{both} the thrust event shape $\tau$ and the transverse momentum $q_T \equiv p_{hT}/z_h$. We assume that neither $z_h$ nor $1-z_h$ is much smaller than one, as we do not resum the additional logarithms that would arise in either limit. For each kinematic region, we present the necessary ingredients to achieve the joint resummation at next-to-next-to-leading logarithmic (NNLL) accuracy. Non-global logarithms occur if the thrust measurement is dominated by the hemisphere that does not contain the observed hadron, since the transverse momentum measurement only probes the radiation in the hemisphere containing the hadron, and we describe how to include the leading NGLs, reducing the accuracy in this limited region of phase space to NLL. The numerical implementation of these factorization formulae, as well as a detailed comparison with Belle data will be presented separately in ref.~\cite{forwardcite}.

While our main focus is on the theoretical framework to describe the Belle measurement~\cite{Seidl:2019jei}, we will also discuss alternative measurements in $e^+e^-$ collisions to constrain TMD FFs. For example, one can measure the hadron transverse momentum relative to the Winner-Take-All (WTA) axis~\cite{Salam:WTAUnpublished,Bertolini:2013iqa}, instead of the thrust axis. If the WTA axis of the hemisphere (jet) containing the hadron is used, this leads to a nonstandard TMD FF~\cite{Neill:2016vbi,Neill:2018wtk}. However, we show that if instead the WTA axis of the other hemisphere is used, this provides direct access to TMD FFs without requiring a cut on additional resolution variables (like thrust) to ensure the dijet limit. 

Interestingly, we note that our formalism can be extended to Deep-Inelastic Scattering (DIS) by crossing the hadron from the final state to the initial state.  We propose a new method for extracting TMD PDFs in DIS, through the angle of the thrust axis with the beam axis in the Breit frame. We develop the factorization theorem for the cross section differential in the Bjorken variable $x$, the photon virtuality $Q$, the DIS thrust (or 1-jettiness), $\tau_{\text{DIS}}$,  and the transverse momentum of the virtual photon w.r.t.~the thrust axis, for the various relevant kinematic regions. The factorization involves the universal (unsubtracted) TMD PDF and soft functions similar to those entering the factorization of the cross section for the Belle measurement. The nonperturbative contribution to the rapidity anomalous dimension is the same as for the CSS kernel~\cite{Collins:1984kg} by consistency of factorization. In some of the kinematic regions, the leading nonperturbative effects associated with soft radiation is the same as in jet-mass measurements in $pp$ collisions~\cite{Stewart:2014nna}.

The outline of our paper is as follows: In \sec{factorization}, we discuss in detail the factorization theorems for the three different kinematic regions. The ingredients needed to achieve the joint resummation at NNLL accuracy are collected in~\sec{ingredients}. In \sec{alternative} we briefly discuss alternative measurements to extract TMD FFs in $e^+e^-$ collisions. In \sec{DIS}, we present results for an analogous measurement using thrust in Deep-Inelastic Scattering (DIS). We conclude and provide an outlook in \sec{conclusions}.

\section{Theoretical framework}
\label{sec:factorization}

In this section we describe the theoretical framework for the $e^+e^- \to hX$ cross section differential in the energy fraction $z_h$ and the transverse momentum $\pmb{q} = \pmb{p}_{h}/z_h$ of the hadron, and the thrust event shape $\tau$. Dividing $\pmb{p}_{h}$ by $z_h$ ensures that $\pmb{q}$ is a partonic variable that can be calculated in perturbation theory for $q_T \equiv |\pmb{q}| \gg \Lambda_{\rm QCD}$. We identify three kinematic regions, depending on the parametric relation between $q_T$ and $\tau$ in \sec{regions}. The corresponding factorization theorems are discussed in secs.~\ref{sec:region1} through \ref{sec:region2}. In \sec{combine} we briefly discuss how to combine these predictions to obtain a matched result for the cross section. This analysis has important parallels with the theoretical framework and predictions for the joint measurement of the transverse momentum of the vector boson and beam thrust in the Drell-Yan process~\cite{Procura:2014cba,Lustermans:2019plv}. A numerical implementation and comparison with experimental data will be presented in ref.~\cite{forwardcite}. 

\subsection{Kinematic bounds}
\label{sec:regions}

First we will argue that the kinematic bound on the phase space is $\sqrt{\t} \gtrsim q_T/Q \gtrsim \t$. To this end, we choose coordinates such that $\hat t = (0,0,1)$, and decompose the 4-momenta $p_i^\mu$ of final-state particles in terms of light-cone coordinates along the thrust axis
\begin{align} \label{eq:lc}
  p_i = p_i^- \frac{n^\mu}{2} + p_i^+ \frac{\bn^\mu}{2} + p_{i\perp}^\mu\,, 
  \quad
  p_i^\pm = p_i^0 \mp p_i^3\,, 
  \quad
  n^\mu = (1,0,0,1)\,,
  \quad
  \bn^\mu = (1,0,0,-1)
\,.\end{align}
This leads to the following expression for the thrust and transverse momentum measurement 
\begin{align} \label{eq:measurement_lc}
  \tau = \frac{1}{Q} \sum_i \min\{p_i^-, p_i^+\}\,, \qquad \pmb{p}_h = -\sum_{\substack{i \in \text{hemi} \\ i \neq h}} \pmb{p}_i \,.
\end{align}
The first expression is obtained by rewriting \eq{thrust} in terms of light-cone coordinates, treating the final-state particles as massless. The second follows from the property of the thrust axis that the total transverse momentum in each separate hemisphere is zero, and the sum on $i$ runs over the hemisphere containing the identified hadron except for the hadron itself, as indicated. 
\begin{figure}[t!]
  \centerline{\includegraphics[width =1.2 \textwidth]{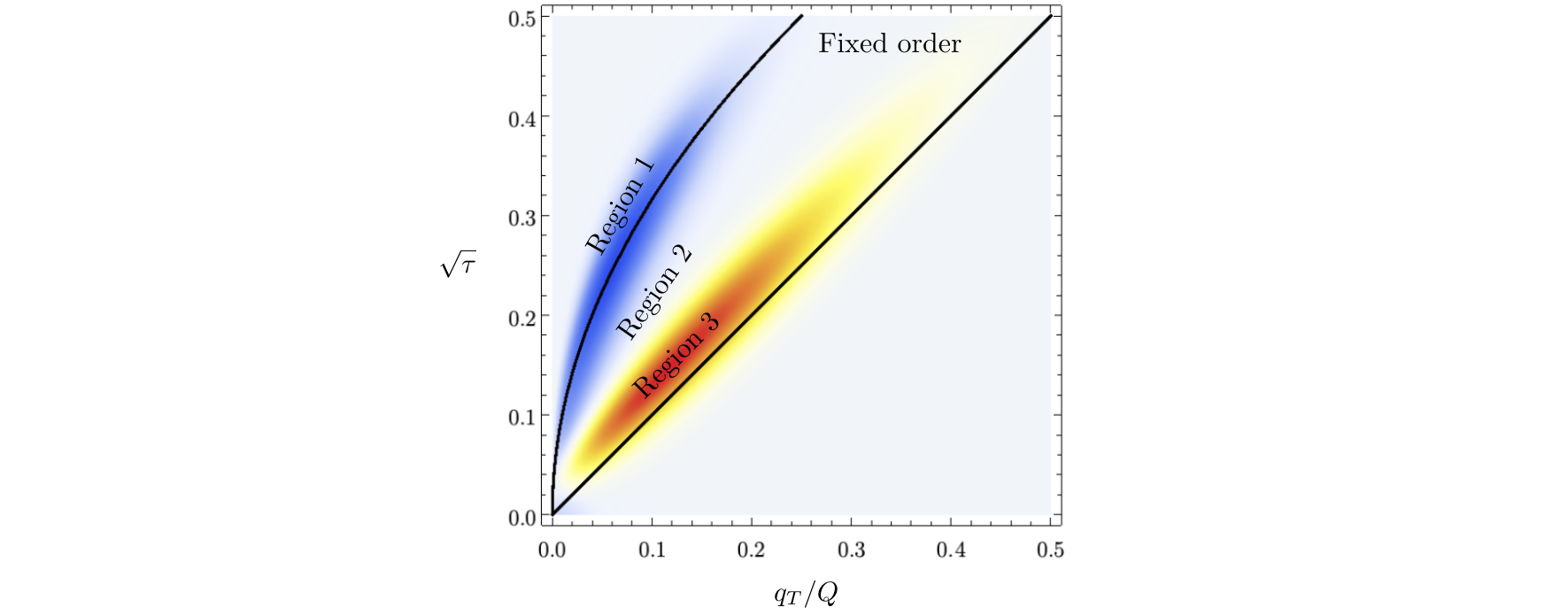}}
  \caption{Illustration of the kinematic regions for the thrust $\tau$ and transverse momentum $q_T/Q$ measurement in $e^+e^-$ collisions, indicated by the different colors.~\label{fig:modes}}
\end{figure}

Employing the (massless) on-shell condition $p_i^- p_i^+ =  \pmb{p}_i^2$,
\begin{align}
  q_T^2 = \frac{1}{z_h^2} \pmb{p}_h^2 = \frac{1}{z_h^2} \biggl(\,\sum_{\substack{i \in \text{hemi} \\ i \neq h}} \pmb{p}_i \biggr)^2  
   \lesssim \sum_i \pmb{p}_i^2 \leq \sum_i  \min\{p_i^-, p_i^+\}Q = \tau Q^2
\,.\end{align}
Here our assumption that $z_h$ is not small entered, in order to conclude that $q_T/Q \lesssim \sqrt{\t}$. 
Conversely,
\begin{align} \label{eq:region1_bnd}
  q_T = \frac{1}{z_h} |\pmb{p}_h| 
  = \frac{1}{z_h} \biggl| \sum_{\substack{i \in \text{hemi} \\ i \neq h}} \pmb{p}_i \biggr|
  \sim \sum_i |\pmb{p}_i| \geq \sum_i  \min\{p_i^-, p_i^+\} = \tau Q\,,
\end{align}
from which we conclude $q_T \gtrsim \tau Q$.
In principle $q_T \ll \tau Q$ is possible: First of all, this could happen because thrust involves a scalar sum and $q_T$ a vector sum. This would require large cancellations in the vector sum, which only occur in a power-suppressed region of phase space. Secondly, \eq{region1_bnd} assumes that the contributions of both hemispheres to thrust are about the same size. However, there can be a large hierarchy between the amount of radiation in each hemisphere, with the more energetic hemisphere dominating the thrust measurement. The $q_T$ measurement is only sensitive to the radiation in the hemisphere containing the observed hadron, see \eq{measurement_lc}, so $q_T \ll \tau Q$ is possible if the observed hadron is in the less energetic hemisphere. This large hierarchy in radiation gives rise to non-global logarithms (NGLs)~\cite{Dasgupta:2001sh}. While there has been substantial progress in resumming NGLs~\cite{Banfi:2002hw,Weigert:2003mm,Schwartz:2014wha,Hagiwara:2015bia,Caron-Huot:2015bja,Larkoski:2015zka,Becher:2016mmh}, there is no practical framework yet to extend this beyond the leading NGLs. Indeed, we will account for these logarithms by starting from the factorization for $q_T \sim \tau Q$ and including them as a correction. The phase-space boundaries are illustrated in \fig{modes} along with the three different kinematic regions for which we develop different factorization formulas below.

Given $\sqrt{\t} \gtrsim q_T/Q \gtrsim \t$, we identify the following three kinematic regions:
\begin{align}
  \text{Region 1:}\quad &\sqrt{\t} \gg q_T/Q \sim \t\,, \nl 
  \text{Region 2:}\quad&\sqrt{\t} \gg q_T/Q \gg \t\,, \nl 
  \text{Region 3:}\quad&\sqrt{\t} \sim q_T/Q \gg \t\,, \nn
\end{align}
in accord with the analysis in ref.~\cite{Procura:2014cba}.
We will always limit ourselves to $q_T/Q \ll1$, since a description of the cross section in terms of TMD FFs is only valid in this case.\footnote{When $q_T/Q \sim 1$, the transverse momentum of the hadron is as large as the hard scale and is generated by the hard scattering. In that case only regular FFs are needed to describe the $z_h$ dependence.} 
Note that this implies $\tau \ll 1$. For each of these regions we now identify the modes in the effective field theory, which are summarized in table~\ref{tab:modes}, and present the factorization theorem that can be used to perform the resummation. 

\begin{table}[t]
\renewcommand{\arraystretch}{1.5}
\centering
   \begin{tabular}{l|ccc} 
     \hline \hline
     Regime: & 1: $\sqrt{\t} \gg q_T/Q \sim \t$ & 2: $\sqrt{\t} \gg q_T/Q \gg \t$ & 3: $\sqrt{\t} \sim q_T/Q \gg \t$ \\
     \hline
     $n$-collinear & $(q_T^2/Q ,Q,q_T)$  & $(q_T^2/Q ,Q,q_T)$ & $Q(\tau,1,\sqrt{\tau})$  \\
     $\bn$-collinear & $Q(1,\tau, \sqrt{\tau})$ & $Q(1,\tau, \sqrt{\tau})$ & $Q(1,\tau,\sqrt{\tau})$ \\
     $n$-collinear-soft & & $(\t Q,  q_T^2 /(\t Q), q_T)$ & \\
     soft & $q_T (1,1,1)$ & $\tau Q(1,1,1)$ & $\tau Q(1,1,1)$ \\
     \hline \hline
   \end{tabular} 
   \caption{The modes (rows) and the corresponding power counting of light-cone components (see \eq{lc}) of momenta for the three kinematic regions (columns).}
    \label{tab:modes}
\end{table}

\subsection{Region 1: $\sqrt{\t} \gg q_T/Q \sim \t$}
\label{sec:region1}

In region 1 (and 3) $q_T/Q$ and $\tau$ are not independent power counting parameters, so an effective theory with only collinear and soft degrees of freedom suffices. We start with a generic power counting for the light-cone components $(p^+, p^-, p_\perp^\mu)$ of their momenta
\begin{align}
  n\text{-collinear:}\quad& p_{n}^{\mu} \sim Q(\lambda_n^2,1, \lambda_n)\,, \nn \\
  \bn\text{-collinear:}\quad& p_{\bn}^{\mu} \sim Q(1,\la_{\bn}^2, \la_{\bn})\,, \nn \\
  \text{soft:}\quad& p_{s}^{\mu} \sim Q(\la_s,\la_s, \la_s)\,.
\end{align}
Taking the hadron along the $n$-direction, \eq{measurement_lc} implies
\begin{align} \label{eq:pc_13}
  \tau \sim \la_n^2 + \la_{\bn}^2 + \la_s \,, \quad q_T/Q \sim \la_n + \la_s
\,.\end{align} 
Combining this with the parametric relation $q_T /Q \sim \t$, we conclude 
\begin{align} \label{eq:pc_1}
\la_{\bn} \sim \sqrt{\tau}\,, \quad \la_n \sim \la_s \sim q_T /Q \sim \t\,,
\end{align}
summarized in the corresponding column in table~\ref{tab:modes}.
Consequently, the $\la_n^2$ contribution to $\tau$ in \eq{pc_13} is power-suppressed and can be dropped. Collinear radiation in the $n$-direction thus only contributes to the transverse momentum measurement of the hadron, collinear radiation in the $\bn$-direction only affects the thrust measurement (this is always the case), while soft radiation contributes to both the thrust measurement and the transverse momentum of the hadron through recoil. 

Throughout this work, we adopt the following definition of the Fourier transform and Laplace transform of a function $f$, as well as their inverse 
\begin{align} \label{eq:FT_LT}
  f(\pmb{b}) &= \int_{-\infty}^{\infty}\! {\rm d} \pmb{q}\; e^{-\img \pmb{b}\cdot \pmb{q}} \, f(\pmb{q})\;, &
  f(\pmb{q}) &= \int_{-\infty}^{\infty}\!  \frac{{\rm d} \pmb{b}}{(2\pi)^2}\; e^{\img \pmb{b}\cdot \pmb{q}} {f}(\pmb{b}) \;,
 \nn \\
  f(u) &= \int_{0}^{\infty}\! \df \tau\; e^{-u\tau}\, f(\tau) \;, &
  f(\tau) &= \frac{1}{2\pi i} \int_{\gamma-i \infty}^{\gamma+i \infty}\! \df u\; e^{u\tau}\, f(u) \;. 
\end{align}
We can then write the factorization theorem for region 1 as 
\begin{align} \label{eq:fact_1}
  \frac{\df \sigma_1}{\df z_h\, \df\pmb{q}\, \df \t } &=   \sum_j  \sigma_{0,j}(Q) \int_{-\infty}^{\infty} \frac{\df \pmb{b}}{(2\pi)^2}  \int_{\gamma-\img \infty}^{\gamma+\img \infty} \frac{\df u}{2\pi \img}\, e^{\img \pmb{b} \cdot \pmb{q} + u\t} H(Q,\mu)\, J\Bigl(\frac{u}{Q^2},\mu\Bigr) \, S \Bigl(\pmb{b},\frac{u}{Q},\mu,\nu\Bigr) 
  \nn \\ & \quad
  \times
  D_{1,j \to h} \Bigl(z_h,\pmb{b},\mu,\frac{\nu}{Q}\Bigr) 
  \biggl[1 + \mathcal{O}\biggl(\tau, \frac{q_T^2}{\tau Q^2}\biggr)\biggr]
\,,\end{align}
where the sum on $j$ runs over (anti-)quark flavors. The variables $\pmb{b}$ and $u$ are the Fourier conjugate of $\pmb{q}$ and  the Laplace conjugate of $\t$, respectively, turning convolutions in the factorization theorem into multiplications. 

We now discuss the various ingredients entering in \eq{fact_1}: The hard function $H$ describes the short-distance $e^+ e^- \to q \bar q$ process, and includes virtual hard corrections.
The $n$-collinear radiation is encoded in the (unsubtracted) TMD FF $D_{1,j \to h}$, which is the object that we aim to extract. The $\bn$-collinear radiation is described by the invariant-mass jet function $J$, which is independent of the (light) quark flavor. The soft function $S$ accounts for the contribution of soft radiation to the thrust event shape and the transverse momentum measurement. Note that only soft radiation in the hemisphere which contains the identified hadron affects $\pmb{b}$, see \eq{measurement_lc}. This does not lead to non-global logarithms, because soft radiation in the other hemisphere is still constrained by the thrust measurement and $q_T/Q \sim \t$. The one-loop expressions for these ingredients are collected in \sec{ingredients}.

The factorization of the cross-section in \eq{fact_1} introduces rapidity divergences which arise from an incomplete factorization of collinear and soft radiation, and are not regulated by dimensional regularization. A range of different regularization procedures have been used to handle these divergences~\cite{Chiu:2007dg,Becher:2010tm,Collins:2011zzd,Chiu:2011qc,Chiu:2012ir,Becher:2011dz,Echevarria:2015usa,Li:2016axz} and in this work we use the $\eta$-regulator~\cite{Chiu:2011qc, Chiu:2012ir}. This leads to a dependence on the rapidity scale $\nu$ in the soft and TMD fragmentation functions. 

The large logarithms in $q_T/Q$ and $\tau$ are resummed by evaluating the ingredients in \eq{fact_1} at their natural renormalization scale $\mu$ and rapidity scale $\nu$ and evolving them to a common $(\mu,\nu)$ scale using the renormalization group equation (RGEs). The  rapidity RGE sums a single logarithm series, and corresponds to the Collin-Soper equation in the CSS approach~\cite{Collins:1984kg}. While we collect explicit expressions for all relevant RGEs in \sec{ingredients}, here we give an instructive example: Consider the hard function $H$, which satisfies
\begin{equation}
\label{eq:example-rge}
    \frac{\df}{\df \ln \mu} H(Q,\mu) = \gamma_H(Q,\mu) H(Q,\mu)\;.
\end{equation}
Evaluating the anomalous dimension $\gamma_H(Q,\mu)$ up to the desired accuracy, we may solve \eq{example-rge} to obtain the evolved hard function,
\begin{equation}
    H(Q,\mu) = H(Q,\mu_H) \exp \lp\int_{\mu_H}^{\mu} \frac{\df \mu}{\mu}  \gamma_H(Q,\mu) \rp
    \stackrel{{\rm LL}}{=} \Bigl(1 - \frac{2\al_s C_F}{\pi} \ln^2 \frac{Q}{\mu_H}\Bigr) \exp\Bigl(- \frac{2\al_s C_F}{\pi} \ln^2 \frac{\mu}{\mu_H}\Bigr)
\end{equation}
The scale $\mu_H$ which sets the initial condition for the RGE is chosen to be $Q$ such that the logarithms in the fixed-order expansion of the hard function are minimized. The RGE then exponentiates (resums) these logarithms, as is clear from the leading logarithmic expression. In terms of the conjugate variables $\pmb{b}$ and $u$, the RGEs are all multiplicative.

The natural renormalization scales of the ingredients in the factorization in \eq{fact_1} are given by 
\begin{align} \label{eq:mu_1}
  \mu_H  \sim Q\,, \quad \mu_J \sim \sqrt{\tau} Q\,, \quad
  \mu_D \sim \mu_S  \sim 1/b_T \sim \tau Q
\,.\end{align}
with $b_T=|\pmb{b}|$ and the corresponding rapidity scales are
\begin{equation}
    \nu_D \sim Q\;,\quad  \nu_S \sim 1/b_T\;.
\end{equation}
While scales can directly be chosen in terms of $\tau$, this leads to spurious divergences for $q_T$~\cite{Frixione:1998dw} that can be avoided by choosing scales in terms of $b_T$. The inverse Fourier transform in \eq{FT_LT} integrates over the nonperturbative region, requiring a prescription to avoid the Landau pole.

We end this section by  addressing the NGLs that arise when $q_T/Q \ll \tau$. Their leading contribution comes from a correlated emission of two gluons, each going into one of the hemispheres. The energy of the emission in the hemisphere containing the hadron is set by $q_T$, whereas the energy of the emission in the other hemisphere is limited by $\tau Q$, leading to
\begin{align}
  S^{\rm NG}\Bigl(\frac{q_T}{\tau Q}\Bigr) = 1 - \frac{\al_s^2 C_F C_A}{12} \ln^2 \Bigl(\frac{q_T}{\tau Q}\Bigr) + \ord{\al_s^3}
\,.\end{align}
The resummed leading NGLs can be included using the standard hemisphere result $S^{\rm NG}$ as an extra factor in \eq{fact_1},
using e.g.~the fit from ref.~\cite{Dasgupta:2001sh}. To obtain the correct \emph{global} logarithms in this case requires one to ``factorize" the soft function 
\begin{align}
  S \Bigl(\pmb{b},\frac{u}{Q},\mu,\nu\Bigr) = S_R\Bigl(\pmb{b},\frac{u}{Q},\mu,\nu\Bigr)  S_L\Bigl(\frac{u}{Q},\mu\Bigr) + \ord{\al_s^2}
\end{align}
where $S_R$ describes emissions into the hemisphere containing the hadron and $S_L$ emissions into the other hemisphere. Taking $\mu_R \sim 1/b_T$ and $\mu_L \sim \tau Q$, which are parametrically different in this NGL regime, is necessary to resum all the global logarithms of $q_T/(\tau Q)$.

\subsection{Region 3: $\sqrt{\t} \sim q_T/Q \gg \t$}
\label{sec:region3}

Next, we discuss the modes and factorization in region 3. Up to \eq{pc_13} the analysis is the same as in region 1. However, the assumption $q_T/Q \sim \sqrt{\t}$ implies 
\begin{align} \label{eq:pc_3}
 \la_{\bn} \sim \la_n \sim \sqrt{\tau} \sim q_T /Q \,, \quad \la_s \sim \tau\,,
\end{align}
instead of \eq{pc_1}. In this case the collinear radiation in the $n$-direction contributes to both the transverse momentum of the hadron and the thrust measurement, while soft radiation only affects thrust, leading to the following factorization theorem~\cite{Jain:2011iu} 
\begin{align} \label{eq:fact_3}
  \frac{d\sigma_3}{\df z_h\, \df\pmb{q}\, \df \t} &=   \sum_j  \sigma_{0,j}(Q)  \int_{\gamma-i \infty}^{\gamma+i\infty} \frac{\df u}{2\pi \img}\, e^{u\t} H(Q,\mu)\, J\Bigl(\frac{u}{Q^2},\mu\Bigr) \, 
  S_{\rm thr}\Bigl(\frac{u}{Q},\mu\Bigr) 
  \mathcal{G}_{j \to h}\Bigl(\frac{u}{Q^2}, z_h,\pmb{q}, \mu\Bigr)
  \nn \\ & \quad \times
    \biggl[1 + \mathcal{O}\biggl(\tau, \frac{q_T^2}{Q^2}, \frac{\tau^2 Q^2}{q_T^2}\biggr)\biggr]
\,.\end{align}
The new ingredients, compared to \eq{fact_1}, are the thrust soft function $S_{\rm thr}$ and the generalized fragmenting jet function (FJF) $\mathcal{G}$, which replaces the TMD FF. From this we conclude that region 3 is not interesting for constraining nonperturbative TMD physics. Indeed, in this case the leading effect of nonperturbative physics arises from the cut on thrust, and the parameter controlling its dominant effect has been fitted from data~\cite{Abbate:2010xh}. 

In \eq{fact_3} only $\mathcal{G}$ contributes to the transverse momentum, so there are no rapidity divergences (nor is there a need to Fourier transform). The large logarithms of $\tau$ and $q_T/Q$ are resummed by evolving all ingredients from their natural scales 
\begin{align}
  \mu_H \sim Q\,, \quad \mu_J \sim \mu_{\mathcal{G}} \sim \sqrt{\tau} Q\,, \quad
   \mu_{S_{\rm thr}}  \sim \tau Q\,.
\end{align}
to a common scale $\mu$, using the RGEs. Explicit expressions for the ingredients in \eq{fact_3} and their anomalous dimensions are collected in \sec{ingredients}.

\subsection{Region 2: $\sqrt{\t} \gg q_T/Q \gg \t$}
\label{sec:region2}

We will now consider region 2, which is the most hierarchical case. If one starts from \eq{pc_13}, the assumption $\sqrt{\t} \gg q_T/Q \gg \t$ implies
\begin{align} \label{eq:pc_2}
\la_{\bn} \sim \sqrt{\t}\,, \quad \la_n \sim q_T /Q\,, \quad \la_s \sim \t\,.
\end{align}
So $\bn$-collinear and soft radiation contribute to thrust, while only $n$-collinear radiation affects the transverse momentum measurement. This does not lead to a consistent factorization theorem. Because $\tau$ and $q_T/Q$ are no longer correlated there is an additional collinear-soft mode, whose power counting is fixed by the fact that it contributes to both $\tau$ and $q_T$~\cite{Procura:2014cba}
\begin{align}
\label{eq:pc_2_cd}
  n\text{-collinear-soft:}\quad& p_{cs}^{\mu} \sim \bigl(\t Q,  q_T^2 /(\t Q), q_T\bigr)
\end{align}

The factorization theorem in region 2 is given by 
\begin{align} \label{eq:fact_2}
  \frac{\df \sigma_2}{\df z_h\, \df\pmb{q}\, \df \t } &=   \sum_j  \sigma_{0,j}(Q) \int_{-\infty}^{\infty} \frac{\df \pmb{b}}{(2\pi)^2}  \int_{\gamma-\img \infty}^{\gamma+\img \infty} \frac{\df u}{2\pi \img}\, e^{\img \pmb{b} \cdot \pmb{q} + u\t} H(Q,\mu)\, J\Bigl(\frac{u}{Q^2},\mu\Bigr)  \, S_{\rm thr}\Bigl(\frac{u}{Q},\mu\Bigr)   
  \nn \\ & \quad
  \times
   \CS\Bigl(\pmb{b},\frac{u}{Q},\mu,\nu\Bigr) \, D_{j \to h}\Bigl(z_h,\pmb{b},\mu,\frac{\nu}{Q}\Bigr)
  \biggl[1 + \mathcal{O}\biggl(\frac{q_T^2}{\tau Q^2}, \frac{\tau^2 Q^2}{q_T^2}\biggr)\biggr]
\,,\end{align}
where the collinear-soft function $\CS$ is the only new ingredient compared to \eqs{fact_1}{fact_3}. Although the arguments of $\CS$ are the same as the soft function $S$ in \eq{fact_1}, its expression differs because the radiation is going into only one hemisphere. Eq.~\eqref{eq:fact_2} contains the \emph{unsubtracted} TMD FF $D_{j \to h}$ and can thus be used for constraining it.

As before, the large logarithms in $q_T/Q$ and $\tau$ are resummed by evaluating the ingredients in \eq{fact_1} at their natural scales and evolving them to a common scale. In this case the natural scales are 
\begin{align}
  \mu_H \sim \nu_D \sim Q\,, \quad \mu_J \sim \sqrt{\tau} Q\,, \quad
  \nu_\CS \sim 1/(b_T^2 \tau Q)\,, \quad
  \mu_D \sim \mu_\CS \sim 1/b_T\,, \quad
  \mu_{S_{\rm thr}} \sim \tau Q\,.
\end{align}
The one-loop expressions and anomalous dimensions for the ingredients entering in \eq{fact_2} are collected in \sec{ingredients}.

\subsection{Matching regions}
\label{sec:combine}

We conclude this section by discussing how to match predictions from the various kinematic regions, described in eqs.~\eqref{eq:fact_1}, \eqref{eq:fact_2} and \eqref{eq:fact_3}. First we note that consistency implies that in region 2 (the most hierarchical case) the factorization theorems for region 1 and 3 must agree with that of region 2. This implies 
\begin{align} \label{eq:consistency}
   S\Bigl(\pmb{b},\frac{u}{Q},\mu,\nu\Bigr)  &=  S_{\rm thr}\Bigl(\frac{u}{Q},\mu\Bigr)   
    \CS\Bigl(\pmb{b},\frac{u}{Q},\mu,\nu\Bigr)
    \Bigl[1 + \mathcal{O}\Bigl(\frac{b_T^2 Q^2}{u^2}\Bigr)\Bigr]
    \,,
    \nn \\
   \mathcal{G}_{j \to h}\Bigl(z_h,\pmb{q}, \frac{u}{Q^2}, \mu\Bigr) &=  
   \int_{-\infty}^{\infty} \frac{\df \pmb{b}}{(2\pi)^2}\,  e^{\img \pmb{b} \cdot \pmb{q}}\,
   \CS\Bigl(\pmb{b},\frac{u}{Q},\mu,\nu\Bigr) 
   D_{j \to h}\Bigl(z_h,\pmb{b},\mu,\frac{\nu}{Q}\Bigr)
   \nn \\ & \quad \times
     \Bigl[1 + \mathcal{O}\Bigl(\frac{q_T^2 u}{Q^2}\Bigr)\Bigr]
\,.\end{align}
It is straightforward to verify this using the expressions in \sec{ingredients}.

While region 2 allows one to resum the most logarithms, since it treats $\tau$ and $q_T/Q$ as uncorrelated, the power corrections in \eq{consistency} grow large as region 1 and 3 are approached. A practical scheme for implementing this was discussed in ref.~\cite{Lustermans:2019plv} for the simultaneous measurement of beam thrust and the transverse momentum of the vector boson in Drell-Yan production. We postpone the full details used in our numerical implementation to ref.~\cite{forwardcite}, but sketch the basic idea here. We start by illustrating the matching of region 1 and fixed-order QCD, 
\begin{align} \label{eq:matching}
\df \sigma
= \df \sigma_1 \bigr|_{\mu_1}
+ \bigl[ \df \sigma_{\rm FO} - \df \sigma_1\bigr]_{\mu_{\rm FO}}
\,,\end{align}
where we suppressed the denominator of the differential cross section for brevity. Here $|\mu_1$ indicates that the scales are chosen according to region $1$, given in \eq{mu_1}. In region 1, the cross section is given by the expected result $\df \sigma_1 \bigr|_{\mu_1}$, since the fixed-order correction is power suppressed. In the fixed-order region, all the resummation is turned off $\mu_1 \to \mu_{\rm FO}$ and $\df \sigma_1 \bigr|_{\mu_{\rm FO}}
+ \bigl[ \df \sigma_{\rm FO} - \df \sigma_1\bigr]_{\mu_{\rm FO}} =  \df \sigma_{\rm FO}|_{\mu_{\rm FO}}$, as required. 

The generalization to the full case is
\begin{align} \label{eq:matching}
\df \sigma
= \df \sigma_2 \bigr|_{\mu_2}
+ \bigl[ \df \sigma_1 - \df \sigma_2 \bigr]_{\mu_1}
+ \bigl[ \df \sigma_3 - \df \sigma_2 \bigr]_{\mu_3}
+ \bigl[ \df \sigma_{\rm FO} - \df \sigma_1 - \df \sigma_3 + \df \sigma_2 \bigr]_{\mu_{\rm FO}}
\,.\end{align}
In region 2, where the most logarithms are resummed, this yields the expected result $\df \sigma_2 \bigr|_{\mu_2}$, as all other terms are power suppressed. As one approaches region 1, the additional resummation from region 2 is turned off by $\mu_2 \to \mu_1$ and the first two terms combine to yield $\si_2|_{\mu_1} + [\si_1 - \si_2]_{\mu_1} = \sigma_1|_{\mu_1}$, while the others encode power corrections. Finally, in the fixed-order region all resummation is turned off, $\mu_i \to \mu_{\rm FO}$, and all terms cancel to leave $\df \sigma_{\rm FO}|_{\mu_{\rm FO}}$. For scales chosen in terms of the impact parameter $b_T$ it is nontrivial to design profiles~\cite{Ligeti:2008ac} that smoothly ensure the correct transitions, e.g.~$\mu_2 \to \mu_1$ as region 1 is approached, since the kinematic regions are not defined in terms of the conjugate variables. In ref.~\cite{Lustermans:2019plv} this was accomplished through a hybrid scale setting.

\section{Fixed-order ingredients}
\label{sec:ingredients}

In this section we collect the one-loop expressions for the ingredients that enter in the factorization theorems in \sec{factorization}, as well as their renormalization group equations. Specifically, we discuss the hard function in \sec{hard}, the invariant-mass jet function in \sec{T_jet}, the thrust soft function in \sec{T_soft}, the thrust-TMD double-differential soft function in \sec{T_TMD_soft}, the thrust-TMD collinear-soft function in \sec{T_TMD_cs}, the unsubtracted TMD FF in \sec{TMD_jet} and the invariant mass-TMD double-differential fragmenting jet function in \sec{T_TMD_jet}.
The anomalous dimensions needed for NNLL resummation are tabulated in appendix~\ref{app:anom}. For convenience we introduce the short-hand notation
\begin{align}
  a_s = \frac{\al_s}{4\pi}\,.
\end{align}

\subsection{Hard function}
\label{sec:hard}

The hard function describes the short-distance $e^+e^- \to q \bar q$ process plus virtual corrections,\footnote{Real corrections to the hard function are not allowed, as they would lead to $q_T \sim Q$.} and 
is thus independent of the kinematic region.
Up to one-loop order it is given by~\cite{Manohar:2003vb,Bauer:2003di}
\begin{equation} \label{eq:H_oneloop}
H(Q, \mu) 
= 1 + a_s \biggl[-2\Ga_0 \ln^2 \Bigl(\frac{\mu}{Q}\Bigr) + \ga_{H\,0} \ln \Bigl(\frac{\mu}{Q}\Bigr) + C_F \Bigl(- 16 + \frac{7\pi^2}{3} \Bigr) \biggr]
\,,\end{equation}
and satisfies the following RGE:
  \begin{align} \label{eq:H_RGE}
\mu \frac{\df}{\df\mu}\, H(Q, \mu) &= \gamma_H(Q, \mu)\, H(Q, \mu) 
\,, \nn \\
\gamma_H(Q, \mu) &=
-4\Gamma_\cusp(a_s) \ln \Bigl(\frac{\mu}{Q}\Bigr) + \gamma_H(a_s)
\,.
\end{align}
Here $\Gamma_\cusp$ is the (quark) cusp anomalous dimension~\cite{Korchemsky:1987wg} and $\gamma_H$ is the non-cusp piece, whose perturbative coefficients $\Ga_n$ and $\ga_{H\,n}$ are tabulated in app.~\ref{app:anom}.

\subsection{Invariant-mass jet function}
\label{sec:T_jet}

The invariant-mass jet function, that accounts for the contribution of the $\bn$-collinear radiation to the thrust measurement in all kinematic regions, is up to one loop given by~\cite{Bauer:2003pi} 
\begin{equation}
J\Bigl(\frac{u}{Q^2},\mu\Bigr) = 1+ a_s \biggl[
\frac12\, \Ga_0 L_J^2 + \frac12\, \ga_{J\,0} L_J + C_F \Bigl(7 -\frac{2\pi^2}{3} \Bigr)   
\biggr]
\,,\end{equation}
rewritten in terms of
\begin{align}
   L_J \equiv \ln\Bigl( \frac{u e^{\ga_E}\, \mu^2}{Q^2}\Bigr)
\,.\end{align}
By using the conjugate variable $u$, its RGE is multiplicative, 
\begin{align} \label{eq:J_RGE}
\mu \frac{\df}{\df\mu}\, J\Bigl(\frac{u}{Q^2},\mu\Bigr) &= \gamma_J\Bigl(\frac{u}{Q^2},\mu\Bigr)\, J\Bigl(\frac{u}{Q^2},\mu\Bigr)
\,, \nn \\
 \gamma_J\Bigl(\frac{u}{Q^2},\mu\Bigr) &=
2\Gamma_\cusp(a_s) L_J + \gamma_J(a_s)
\,.
\end{align}
The coefficients $\Ga_n$ and $\ga_{J,n}$ of the anomalous dimension are given in app.~\ref{app:anom}.

\subsection{Thrust soft function}
\label{sec:T_soft}

The one-loop thrust soft function, which describes the contribution to thrust from soft radiation in region 2 and 3, is given by
\begin{equation}
S_{\rm thr}\Bigl(\frac{u}{Q},\mu\Bigr) = 1+ a_s \bigl[  -2 \Ga_0 L_S^2 - C_F \pi^2 \bigr]
\,,\end{equation}
up to one-loop order, where 
\begin{align} \label{eq:LS}
   L_S \equiv \ln\Bigl( \frac{u e^{\ga_E}\, \mu}{Q}\Bigr)
\,.\end{align}
It's multiplicative RGE has anomalous dimension
\begin{align} \label{eq:S_RGE}
\gamma_{S_{\rm thr}}\Bigl(\frac{u}{Q},\mu\Bigr) &=
-4\Gamma_\cusp(a_s) L_S + \gamma_{S_{\rm thr}}(a_s)
\,.
\end{align}
The non-cusp anomalous dimension satisfies $\gamma_{S_{\rm thr}}(a_s) = - \ga_H(a_s) - 2 \ga_J(a_s)$, due to consistency of the thrust factorization theorem.

\subsection{Thrust-TMD soft function}
\label{sec:T_TMD_soft}

The double-differential soft function $S$ describes the effect of soft radiation in regime 1, accounting for both the contribution to the thrust and transverse momentum measurement. 
At one-loop order, only a single soft gluon is emitted, which goes either into the hemisphere that contains the identified hadron or the other hemisphere. It only affects the TMD measurement if it is emitted in the same hemisphere as the identified hadron, due to \eq{measurement_lc}. Consequently, at this order we can obtain $S$ from two known soft functions, 
\begin{equation}
  S\Bigl(\pmb{b},\frac{u}{Q},\mu,\nu\Bigr) = \frac{1}{2} S_{\rm thr}\Bigl(\frac{u}{Q},\mu\Bigr) + \frac12  S_{pp}\Bigl(\pmb{b},\frac{u}{Q},\mu,\nu\Bigr) + \mathcal{O}(a_s^2)
\,,\end{equation}
where the thrust soft function $S$ was given in \sec{T_soft}. The double-differential $S_{pp}$ was calculated in ref.~\cite{Procura:2014cba} and can be conveniently written in terms of the TMD soft function and a remainder,
\begin{equation}
S_{pp}\Bigl(\pmb{b},\frac{u}{Q},\mu,\nu\Bigr) = S(\pmb{b},\mu,\nu) + \Delta S^{(1)}\Bigl(\pmb{b},\frac{u}{Q},\mu\Bigr) + \mathcal{O}(a_s^2)
\,.\end{equation}
The TMD soft function is given by~\cite{Chiu:2012ir}
\begin{equation}
S(\pmb{b},\mu,\nu) = 1+ a_s \biggl\{ \Ga_0\biggl[ -\frac12  L_b^2 - 2 L_b \ln \lp \frac{\nu}{\mu}\rp\biggr] - C_F \frac{\pi^2}{3}\biggr\}
\,.\end{equation}
in terms of
\begin{equation}
   L_b \equiv 2 \ln \Bigl( \frac{b_T \mu}{2 e^{-\ga_E}} \Bigr)
\,.\end{equation}
For the remainder $\De S^{(1)}$ we use the following result in our numerical analysis~\cite{Lustermans:2019plv}
\begin{align}
\Delta S^{(1)}_c (\bmat{b},k_c,\mu) &= \int_0^{k_c} \df k \, \frac{1}{2\pi \img} \int_{\gamma-\img \infty}^{\gamma+\img \infty } \df u\, \exp\Bigl(\frac{u k}{Q} \Bigr)\, \Delta S^{(1)}\Bigl(\pmb{b},\frac{u}{Q},\mu\Bigr)
\nn \\ &
=  4 a_s C_F \Bigl[ g (b_T k_c/2) - 2\ln^2 \lp \frac{b_T k_c e^{\gamma_E}}{2}\rp \Bigr]
\end{align}
which is the cumulative result for $k \leq k_c$, with
\begin{equation}
g(x) \equiv x^2\ {}_3F_4 (1,1,1;2,2,2,2;- x^2 ) 
\,.\end{equation}
The double-differential soft function $S$ satisfies the following RGE 
\begin{align} \label{eq:SS_RGE}
  \mu\frac{\df }{\df\mu}  S\Bigl(\pmb{b},\frac{u}{Q},\mu,\nu\Bigr) &=   \gamma_S \Bigl(\frac{u}{Q},\mu,\nu\Bigr)\, S\Bigl(\pmb{b},\frac{u}{Q},\mu,\nu\Bigr)
\,, \nn \\
\nu \frac{\df}{\df\nu}\,  S\Bigl(\pmb{b},\frac{u}{Q},\mu,\nu\Bigr) &= \frac12 \ga_\nu(\pmb{b},\mu)\, S\Bigl(\pmb{b},\frac{u}{Q},\mu,\nu\Bigr)
\,, \nn \\
  \gamma_S \Bigl(\frac{u}{Q},\mu,\nu\Bigr) &= -2 \Ga_\cusp(a_s) \Bigl[L_S + \ln \lp \frac{\nu}{\mu} \rp\Bigr] + \ga_S(a_s)
  \,, \nn \\
    \gamma_{\nu} (\pmb{b},\mu) &= - 2 \Gamma_{\cusp} L_b + \ga_\nu(a_s)
\end{align}
where $\ga_\nu$ is the universal rapidity anomalous dimension and $L_S$ is defined in \eq{LS}. The non-cusp anomalous dimension $\ga_S$ vanishes up to two-loop order 
(for the exponential regulator~\cite{Li:2016axz} this is true to all orders).

\subsection{Thrust-TMD collinear-soft function}
\label{sec:T_TMD_cs}

The double differential collinear-soft function describes $n$-collinear radiation in kinematic region 2, encoding both the contribution to the TMD and thrust measurement. Up to one-loop order it is given by~\cite{Procura:2014cba}
\begin{equation}
  \CS \Bigl(\pmb{b},\frac{u}{Q},\mu,\nu\Bigr) = 1 +  a_s \biggl\{\Ga_0  \biggl[- \frac12  L_b^2 + L_b \biggl(L_S +   
  \ln \Bigl(\frac{\mu}{\nu}\Bigr) \biggr) \biggr]- C_F\, \frac{\pi^2}{3} \biggr\}
\,.\end{equation}
It satisfies the following RGE in $\mu$ and $\nu$
\begin{align} \label{eq:CS_RGE}
\mu \frac{\df}{\df\mu}\, \CS\Bigl(\pmb{b},\frac{u}{Q},\mu,\nu\Bigr) &= \gamma_\CS \Bigl(\frac{u}{Q},\mu,\nu\Bigr)\, \CS\Bigl(\pmb{b},\frac{u}{Q},\mu,\nu\Bigr)
\,, \nn \\
\nu \frac{\df}{\df\nu}\, \CS\Bigl(\pmb{b},\frac{u}{Q},\mu,\nu\Bigr) &= \frac12 \ga_\nu(\pmb{b},\mu)\, \CS\Bigl(\pmb{b},\frac{u}{Q},\mu,\nu\Bigr)
\,, \nn \\
\gamma_\CS\Bigl(\frac{u}{Q},\mu,\nu\Bigr) &=
2\Gamma_\cusp(a_s) \Bigl[L_S +   
  \ln \Bigl(\frac{\mu}{\nu}\Bigr) \Bigr]  + \gamma_\CS(a_s)
\,,\end{align}
where $\ga_\nu$ is the rapidity anomalous dimension in \eq{SS_RGE} and the perturbative expansion of $\ga_\CS$ is given in app.~\ref{app:anom}.

\subsection{Unsubtracted TMD fragmentation function}
\label{sec:TMD_jet}

The unsubtracted TMD FFs $ D_{i/h}(z,\pmb{b},\mu,\nu/Q)$ were introduced in SCET in refs.~\cite{Echevarria:2014rua,Bain:2016rrv}. 
In the perturbative regime, they can be matched onto collinear FFs~\cite{Aybat:2011zv,Echevarria:2014rua,Bain:2016rrv, Kang:2017glf} 
\begin{equation}
  \label{eq:FF_match}
  D_{q\to h}\Bigl(z,\pmb{b},\mu,\frac{\nu}{Q}\Bigr) = \int_{z}^1 \frac{\df x}{x}\; \mathcal{J}_{q\to j}\Bigl(x, \pmb{q},\mu,\frac{\nu}{Q}\Bigr) \;D_{j \to h}\lp \frac{z}{x},\mu  \rp \Bigl[1 + \mathcal{O}(\Lambda_{\text{QCD}}^2 b_T^2)\Bigr]
\end{equation}
The renormalized matching coefficients are given by~\cite{Bain:2016rrv}
\begin{align}
  \mathcal{J}_{q\to j}\Bigl(x,\pmb{b},\mu,\frac{\nu}{Q}\Bigr) &= \delta_{qj}\delta(1-x)
   \\ & \quad
  + a_s  \biggl\{ \biggl[ \delta_{qj} \Bigl(\Ga_0 \ln\lp\frac{\nu}{Q}\rp + \frac12 \ga_{D,0} \Bigr) \delta(1-x) -2 C_F P_{jq}(x)\biggr] L_b
  +c_{qj}(x)
  \biggr\},
\nn\end{align}
up to one-loop order, where the splitting functions $P_{jq}$ and finite terms are given by
\begin{align} \label{eq:P_and_c}
  {P}_{qq}(x)&= \Bigl(\frac{1+x^2}{1-x}\Bigr)_+\,,  \qquad
  {P}_{gq}(x)=\frac{1+(1-x)^2}{x}\,,  \nn \\
  {c}_{qq}(x)&=2 C_F (1-x)\,, \qquad
  {c}_{qg}(x)=2 C_F x\,. 
\end{align}
The corresponding $\mu$ and $\nu$ renormalization group equations are 
\begin{align} \label{eq:D_RGE}
\mu \frac{\df}{\df\mu}\, D\Bigl(z,\pmb{b},\mu,\frac{\nu}{Q}\Bigr) &= \gamma_D\Bigl(\mu,\frac{\nu}{Q}\Bigr)\, D\Bigl(z,\pmb{b},\mu,\frac{\nu}{Q}\Bigr)
\,, \nn \\
\nu \frac{\df}{\df\nu}\, D\Bigl(z,\pmb{b},\mu,\frac{\nu}{Q}\Bigr) &= -\frac12 \ga_\nu(\pmb{b},\mu)\, D\Bigl(z,\pmb{b},\mu,\frac{\nu}{Q}\Bigr)
\,, \nn \\
\gamma_D\Bigl(\mu,\frac{\nu}{Q}\Bigr)&= 2 \Ga_{\cusp}(a_s) \ln\lp \frac{\nu}{Q}  \rp + \ga_D(a_s)
\,.\end{align}

\subsection{Invariant mass-TMD fragmenting jet function}
\label{sec:T_TMD_jet}

The fragmenting jet function encodes the effects of $n$-collinear radiation in region 3, contributing to both the thrust and TMD measurement. In the perturbative regime this can be matched onto collinear fragmentation functions (similar to the matching for TMD FFs in \eq{FF_match})
\begin{equation} \label{eq:FJF_match}
  \mathcal{G}_{q\to h}\Bigl(z,\pmb{q}, \frac{u}{Q^2},\mu\Bigr) = \int_{z}^1 \frac{\df x}{x}\; \mathcal{J}_{q\to j}\Bigl(x, \pmb{q}, \frac{u}{Q^2},\mu\Bigr) \;D_{j \to h}\lp \frac{z}{x},\mu  \rp
\,.\end{equation}
Up to one-loop order, the matching coefficients are given by~\cite{Jain:2011iu} 
\begin{align}
  \mathcal{J}_{q\to j}\Bigl(x, \pmb{q},\frac{u}{Q^2},\mu\Bigr) &= \de_{qj} \de^{(2)}(\pmb{q}) \delta(1\!-\!x) + a_s  \biggl\{
   \de_{qj}  \biggl[ \frac12  \Ga_0  L_J^2 +
   \frac12\, \ga_{J\,0} L_J \biggr]  
    \delta^{(2)}(\pmb{q}) \delta(1\!-\!x)  
  \nn \\  & \quad 
  + \frac{2C_F}{\pi} \frac{1}{(1-x)\mu^2} \Bigl(\frac{(1-x)\mu^2}{q_T^2}\Bigr)_+ \!\! \exp\Bigl( - \frac{x}{1-x} \frac{u\, q_T^2}{Q^2}  \Bigr) P_{jq}(x) 
  \nn \\ & \quad
  + \bigl[4C_F P_{jq}(x) \ln x + d_{qj}(x) \bigr] \de^{(2)}(\pmb{q}) \biggr\}
\,.\end{align}
where the splitting functions are given in \eq{P_and_c} and 
\begin{align}
  {d}_{qq}(x)&=2 C_F \Bigl[(1+x^2) \Bigl(\frac{\ln(1-x)}{1-x}\Bigr)_+ + 1-x \Bigr]\,, \qquad
  {d}_{qg}(x)=2 C_F \bigl[P_{gq}(x) \ln(1-x) + x\bigr]\,. 
\end{align}
The corresponding anomalous dimension is equal to the jet function anomalous dimension $\gamma_{J}$ in \eq{J_RGE}, as required by consistency of the factorization theorem.

\section{Alternative methods to extract TMD FFs}
\label{sec:alternative}

While the main focus of this paper has been the extraction of TMD FFs from the transverse momenta of hadrons with respect to the thrust axis, we now comment on some different measurements that can be used to extract the TMD FFs. 

First of all, one can consider the transverse momentum of a hadron with respect to the jet axis in $pp$ collisions~\cite{Bain:2016rrv, Kang:2017glf}, which also provides access to fragmentation initiated by gluons. To avoid  jet boundary effects that may distort the transverse momentum distribution, $q_T \ll p_T R$ is required, where $p_T$ is the jet transverse momentum and $R$ the radius of the jet. Because the jet axis points along the jet momentum, the sum of momenta transverse to the \emph{jet axis} is zero. Consequently, soft radiation inside the jet affects the transverse momentum measurement through its recoil of the jet axis, while soft radiation outside the jet is unconstrained. This leads to non-global logarithms that currently limit the theoretical accuracy to NLL. 

\begin{figure}[t!]
  \centerline{\includegraphics[width = \textwidth]{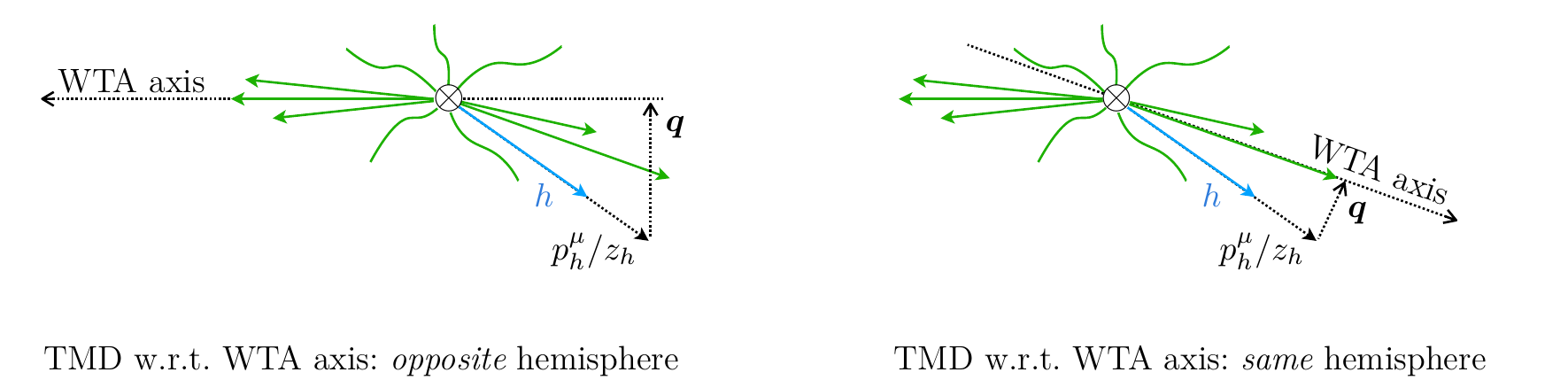}}
  \caption{The transverse momentum of an identified hadron with respect to the WTA axis in the opposite hemisphere (left) or same hemisphere (right).~\label{fig:WTA-setups}}
\end{figure}

Alternatively, one can consider the transverse momentum of a hadron in a jet with respect to a recoil-free axis~\cite{Neill:2016vbi}, such as the Winner-Take-All (WTA) axis~\cite{Salam:WTAUnpublished,Bertolini:2013iqa}. This is illustrated in the right panel of \fig{WTA-setups}. This removes all sensitivity to soft radiation, but the resulting transverse momentum distributions are not the standard TMD FFs. Indeed, rather than Sudakov double logarithms of the transverse momentum, the corresponding cross section involves single logarithms that are resummed by a modified DGLAP evolution~\cite{Neill:2016vbi,Neill:2018wtk}.

Within the context of $e^+e^-$ collisions, there is another interesting alternative: One can determine the WTA axis in one hemisphere and consider the transverse momentum of a hadron in the \emph{other} hemisphere with respect to this axis, shown in the left panel of \fig{WTA-setups}.\footnote{One can either identify the hemispheres using e.g.~exclusive $k_T$ or taking the hemisphere defined by the thrust axis. The difference between these choices is power suppressed for small transverse momenta.} The resulting cross section has several advantages: There is no additional cut on thrust needed to ensure the two-jet limit, as this is already taken care of by requiring that the transverse momentum $q_T \ll Q$. There are no NGLs because soft radiation is treated the same, independent of the hemisphere it is emitted into, with only its total recoil affecting $q_T$. The corresponding cross section is given by
\begin{align} 
  \frac{\df \sigma_{\rm WTA}}{\df z_h\, \df\pmb{q}} &= \sigma_0 \int_{-\infty}^{\infty} \frac{\df \pmb{b}}{(2\pi)^2}\,  e^{\img \pmb{b} \cdot \pmb{q}} H(Q,\mu)  \mathscr{J} \Bigl(\pmb{b},\mu,\frac{\nu}{Q}\Bigr)   S(\pmb{b},\mu,\nu) 
    \sum_j D_{1,j \to h} \Bigl(z_h,\pmb{b},\mu,\frac{\nu}{Q}\Bigr) 
  \nn \\ & \quad
  \times
  \biggl[1 + \mathcal{O}\biggl(\frac{q_T^2}{Q^2}\biggr)\biggr]
\,,\end{align}
where the jet function $\mathscr{J}$ encodes the offset of the winner-take-all axis and the collinear quark that initiates the jet~\cite{Gutierrez-Reyes:2018qez,Gutierrez-Reyes:2019vbx}. Note that this is identical to the factorization formulae for dihadron fragmentation, with one of the TMD FFs replaced by a TMD jet function. The observable proposed here describes the transverse momentum of a hadron with respect to a jet in the opposite direction, whose axis is found using the WTA scheme. It provides a unique crosscheck between dihadron $hh$ and dijet $JJ$ production (both with WTA axis) in $e^+e^-$ collisions. Interestingly, this same TMD jet function also appears in studies of the TMD PDF in SIDIS using jets~\cite{Gutierrez-Reyes:2018qez,Gutierrez-Reyes:2019vbx}, and the azimuthal angular decorrelation in vector boson plus jet production in $pp$ collisions~\cite{Chien:2020hzh}.

\section{Extracting TMD PDFs in DIS using the thrust axis}
\label{sec:DIS}

\begin{figure}[t!]
  \centerline{\includegraphics[width = \textwidth]{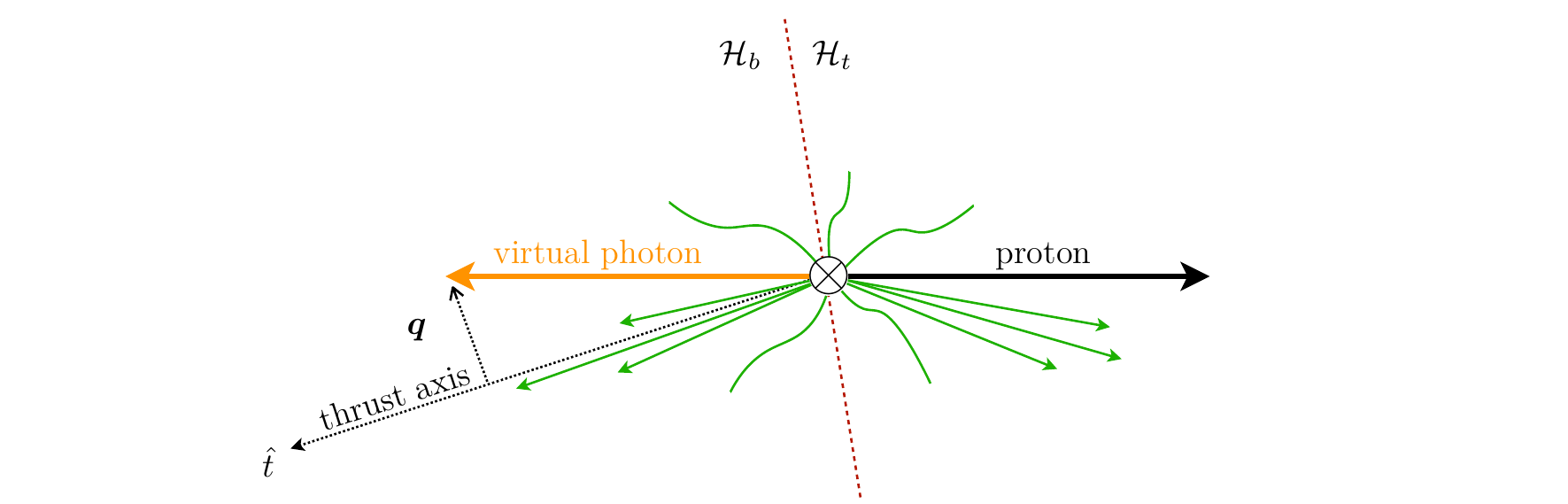}}
  \caption{Illustration of the measurement of the virtual photon transverse momentum w.r.t. the thrust axis, $\bmat{q}$, in the Breit frame.~\label{fig:measurementDIS} }
\end{figure}

In this paper we focused on the formalism for describing the double differential spectrum of $\bmat{q}$ and $\tau$ in electron-positron collisions, motivated by its recent measurement by the Belle collaboration. However, our formalism can also be extended to deep-inelastic-scattering (DIS), offering a unique approach to extract TMD PDFs, as we now discuss.

Consider DIS in the Breit frame, for which the virtual photon momentum is given by:
\begin{equation}
    q^{\mu}  = Q/2( \bn^{\mu} - n^{\mu} ) = Q (0,0,0,-1)\,,
\end{equation}
where $n^{\mu} \equiv (1,0,0,1)$ and $\bn^{\mu} \equiv (1,0,0,-1)$.
Up to mass corrections, the proton momentum can be written as
\begin{equation}
    P^{\mu} \simeq Q/(2x) n^{\mu} = Q/(2x) (1,0,0,1)\,,
\end{equation} 
where $x \equiv Q^2 / (2 \,q\cdot P)$ is the Bjorken variable. At Born level, the struck quark back-scatters against the photon with momentum ($x\simeq \xi$) 
\begin{equation}
    p_q^{\mu} = \xi P^{\mu} + q^{\mu} \simeq (Q/2) \bn^{\mu}\,.
\end{equation}
The struck-quark fragments and produces a jet-like structure which points (close) to the opposite of the beam direction. 

We propose the measurement of the photon transverse momentum  $\bmat{q}$ in the Breit frame, with respect to the thrust axis $\hat{t}$, when a cut on the DIS-thrust event shape is imposed, $\tau_{\text{DIS}} < \tau_c$. The thrust axis and DIS event shape we use are defined by
\begin{equation}
    \tau_{\text{DIS}} = \min_{\hat{t}} \frac{1}{Q}\sum_i \min \bigl\{ n\cdot p_i, n_t \cdot p_i \bigr\} \,,
\end{equation}
where  $n_t^{\mu} \equiv (1,\hat{t}\,)$ is determined through minimization. Up to proton mass corrections, this definition is closely related to the 1-jettiness for DIS of ref.~\cite{Kang:2012zr} and $\tau_1^a$' of ref.~\cite{Kang:2013nha}.
In principle one can use a Lorentz invariant definition, as in ref.~\cite{Kang:2013nha}, but we find it convenient to give the simpler definition for the Breit frame. 
For alternative definitions of DIS-thrust see also refs.~\cite{Antonelli:1999kx,Dasgupta:2002dc, Dasgupta:2003iq}. To keep the notation compact, we drop the subscript DIS in the following, simply using $\tau$.

This measurement is related to the transverse momentum of a hadron in $e^+e^-$ collisions by crossing the outgoing hadron to an incoming proton, thereby replacing a TMD FF by a TMD PDF. In particular, the kinematic regions and modes discussed in \sec{factorization} are the same. We now discuss the factorization using \fig{measurementDIS}, where we have separated the phase space into the hemispheres $\mathcal{H}_b$ and $\mathcal{H}_t$, defined by
\begin{align}
   i \in \mathcal{H}_b &:\quad n\cdot p_i > n_{t}\cdot p_i\,, \nn \\
   i \in \mathcal{H}_t &:\quad n \cdot p_i < n_{t}\cdot p_i\,,
\end{align}
 where $p_i^{\mu}$ is the (massless) momentum of particle $i$. 
 
\begin{figure}[t!]
  \centerline{\includegraphics[width = \textwidth]{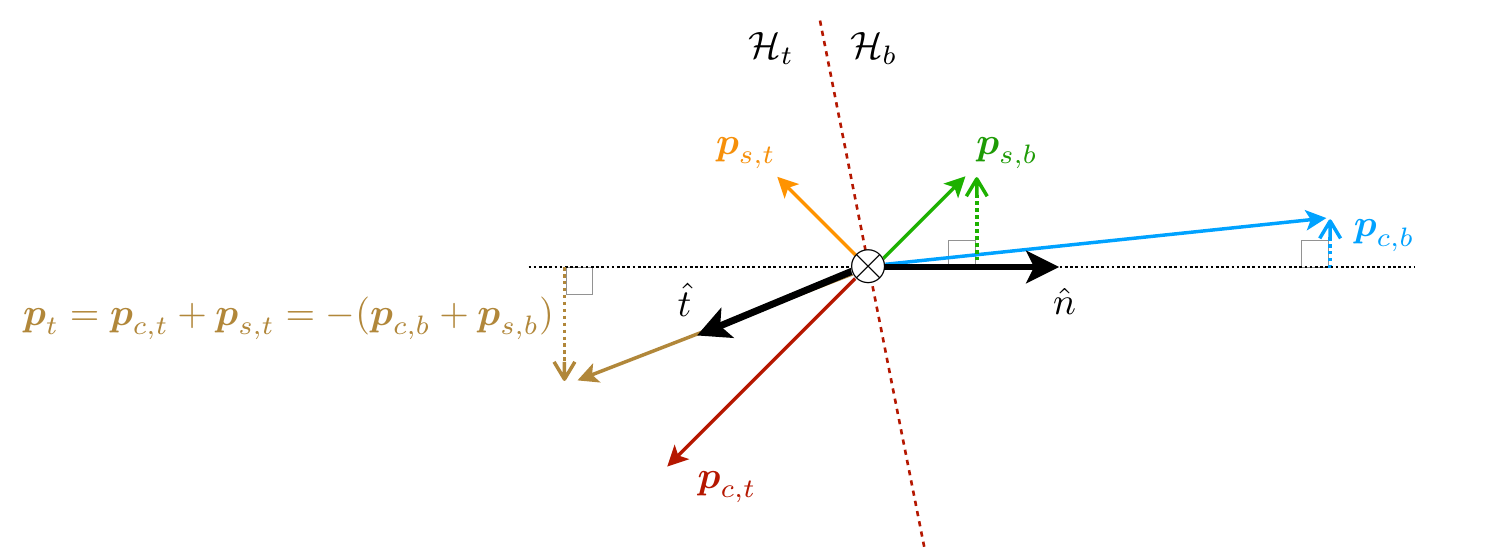}}
  \caption{Momentum conservation in region 1.~\label{fig:kinematicsDIS} }
\end{figure}

In Region 1, which as before satisfies $\t \sim |\bmat{q}|/ Q \ll \sqrt{\t}$, soft radiation contributes to both the thrust measurement and the TMD measurement. Initial-state collinear radiation along the direction of the proton will contribute to $\mathcal{H}_b$ with momentum $p_{c,b}^{\mu}$, and final-state collinear radiation from the struck quark will contribute to $\mathcal{H}_t$, with momentum $p_{c,t}^{\mu}$. Soft radiation will be emitted in both hemispheres, and we denote the momentum of its contribution to $\mathcal{H}_b$ and $\mathcal{H}_t$ with $p_{s,b}^{\mu}$ and $p_{s,t}^{\mu}$, respectively. The transverse momentum $\bmat{q}$ of the photon with respect to the thrust axis, is directly related to the total transverse momentum $\bmat{p}_t$ of radiation in hemisphere $\mathcal{H}_t$ w.r.t. proton/photon axis, through $\bmat{q} = -2 \bmat{p}_t$.  By momentum conservation  $\bmat{p}_t = - (\bmat{p}_{c,b}+ \bmat{p}_{s,b})$, showing explicitly that for the transverse momentum measurement only contributions from the initial-state collinear radiation and soft radiation in $\mathcal{H}_b$ are relevant. On the other hand, the thrust measurement receives contributions from the final-state collinear radiation and soft radiation from both hemispheres. The kinematics and momentum conservation from contribution of the various modes is illustrated in \fig{kinematicsDIS}. This leads to the following factorization theorem, 
 \begin{align} \label{eq:DIS_fact_1}
  \frac{\df \sigma_1}{\df x \,\df Q^2 \, \df\bmat{p}_t\, \df \t } &=  \sum_j \sigma_{0,j}(x,Q) \int_{-\infty}^{\infty} \frac{\df \pmb{b}}{(2\pi)^2}  \int_{\gamma-\img \infty}^{\gamma+\img \infty} \frac{\df u}{2\pi \img}\, e^{\img \pmb{b} \cdot \pmb{p}_t + u\t} H(Q,\mu)\, J\Bigl(\frac{u}{Q^2},\mu\Bigr)\, 
  \nn \\ & \quad
  \times
  S \Bigl(\pmb{b},\frac{u}{Q},\mu,\nu\Bigr) \,
  F_{1,j} \lp x,\pmb{b},\mu,\frac{\nu}{Q}\rp 
  \lb 1 + \mathcal{O}\lp \tau, \frac{q_T^2}{\tau Q^2}\rp \rb
\,,\end{align}
where $F_{1,j}$ is the unsubtracted unpolarized  quark TMD PDF. We leave the effect of polarized incoming protons for future investigations. The soft function $S$, and in general the global soft functions that appear in this section, are related to the corresponding matrix elements in the $e^+e^-$ factorization by exchanging an outgoing Wilson line for an incoming one. For compactness of the notation, we avoid the use of new symbols or superscripts.

To obtain the factorization for the other kinematic regions involves the same steps. In region 2, the collinear-soft initial-state radiation also contributes to  $\bmat{p}_t$ while the soft contribution is power suppressed:
$\bmat{p}_t  = -(  \bmat{p}_{c,b}+ \bmat{p}_{cs,b})$.
The corresponding factorization theorem is
\begin{align} \label{eq:DIS_fact_2}
  \frac{\df \sigma_2}{\df x \,\df Q^2 \, \df\bmat{p}_t\, \df \t } &= \sum_j \sigma_{0,j}(x,Q) \int_{-\infty}^{\infty} \frac{\df \pmb{b}}{(2\pi)^2}  \int_{\gamma-\img \infty}^{\gamma+\img \infty} \frac{\df u}{2\pi \img}\, e^{\img \pmb{b} \cdot \pmb{p}_t + u\t} H(Q,\mu)\, J\Bigl(\frac{u}{Q^2},\mu\Bigr)\,   S_{\rm thr}\Bigl(\frac{u}{Q},\mu\Bigr)   
  \nn \\ & \quad
  \times
  \CS\Bigl(\pmb{b},\frac{u}{Q},\mu,\nu\Bigr) 
  F_{1, j}\lp x ,\pmb{b},\mu,\frac{\nu}{Q}\rp
  \lb 1 + \mathcal{O}\lp\frac{q_T^2}{\tau Q^2}, \frac{\tau^2 Q^2}{q_T^2}\rp\rb
\,.\end{align}
Finally, for region 3 the power counting in \eq{pc_3} implies $\bmat{p}_t =-  \bmat{p}_{c,b}\,$.
The corresponding factorization theorem is,
\begin{align} \label{eq:DIS_fact_3}
  \frac{\df\sigma_3}{\df x\, \df Q^2\,\df\pmb{p}_t\, \df \t} &= \sum_j \sigma_{0,j} (x,Q)  \int_{\gamma-i \infty}^{\gamma+i\infty} \frac{\df u}{2\pi \img}\, e^{u\t} H(Q,\mu)\, J\Bigl(\frac{u}{Q^2},\mu\Bigr)\,  
  S_{\rm thr} \Bigl(\frac{u}{Q},\mu\Bigr) 
   \mathcal{B}_{j}\Bigl(\frac{u}{Q^2}, x,\pmb{p}_t, \mu\Bigr)
  \nn \\ & \quad \times
    \lb1 + \mathcal{O}\lp\tau, \frac{q_T^2}{Q^2}, \frac{\tau^2 Q^2}{q_T^2}\rp\rb
\,.\end{align}
where $\mathcal{B}_{j}$ is the Laplace transform of the double differential (in thrust and transverse momentum) beam function~\cite{Jain:2011iu,Gaunt:2014xxa}. 

\section{Conclusions}
\label{sec:conclusions}

We have presented a factorization framework for the transverse momentum of hadrons measured in $e^+e^-$ collisions relative to the thrust axis. We identified three kinematic regions depending on the relative scaling of the transverse momentum of the hadrons and the thrust event shape. One of these kinematic regions was studied before in~\cite{Jain:2011iu}, but is not relevant for extracting TMD FFs, since they do not appear in the corresponding factorization theorem. For each of these regions, we developed the necessary factorization formulas using Soft Collinear Effective Theory. We presented analytical results for the different functions that appear in the factorization, which allow for the joint resummation of logarithms of the transverse momentum and thrust at next-to-next-to-leading logarithmic accuracy. Our studies were motivated by recent measurements of the Belle Collaboration differential in the transverse momentum of hadrons and cumulative in thrust. We will present numerical studies for this process and a detailed comparison to the experimental data in ref.~\cite{forwardcite}. Using this new data set it will be possible to (more) precisely constrain transverse momentum dependent fragmentation functions (TMD FFs).

In addition, we proposed to constrain TMD FFs by identifying the winner-take-all axis in $e^+e^-$ collisions in one hemisphere and measuring the transverse momentum of hadrons in the opposite hemisphere with respect to this axis. This process is simpler than the double differential cross section described above, which is the main focus of this work, and it is free of non-global logarithms. Furthermore, we proposed to measure the transverse momentum of the virtual photon in deep-inelastic scattering with respect to the thrust axis in the Breit frame. This process can be thought of as a crossed version of the double differential $e^+e^-$ cross section described above. This DIS measurement opens up a new avenue to access transverse momentum dependent parton distribution functions (TMD PDFs), which is independent of TMD FFs and jet clustering effects. We presented the relevant factorization formulas and we expect that this measurement will be a powerful probe of TMD PDFs, especially at the future Electron-Ion Collider.

\subsection*{Acknowledgments} 

We thank Johannes Michel, Ralf Seidl and Anselm Vossen for discussions, Johannes Michel and Zhong-Bo Kang for feedback on the manuscript, and Zhong-Bo Kang for collaborating in early stages of this project. 
YM is supported by the European Union’s Horizon 2020 research and innovation program under the Marie Sk\l{}odowska-Curie grant agreement
No. 754496-FELLINI. 
FR is supported by LDRD funding from Berkeley Lab provided by the U.S. Department of Energy under Contract No. DE-AC02-05CH11231.
WW is supported by the ERC grant ERC-STG-2015-67732 and the D-ITP consortium, a program of NWO funded by the Dutch Ministry of Education, Culture and Science (OCW).  

\appendix

\section{Anomalous dimensions}
\label{app:anom}

Expanding the anomalous dimensions in powers of $a_s = \alpha_s/(4\pi)$,
\begin{align}
\Gamma_\cusp(a_s) = \sum_{n=0}^\infty \Gamma_na_s^{n+1}
\,, \quad
\gamma_X(a_s) = \sum_{n=0}^\infty \gamma_{X,n} a_s^{n+1}
\,,\end{align}
the coefficients that are needed up to NNLL order are given by 
\begin{align}
\Gamma_0 &= 4 C_F
\,,\nn\\
\Gamma_1 &= 4 C_F \Bigl[C_A\,\Bigl( \frac{67}{9} -\frac{\pi^2}{3} \Bigr)  -
   \frac{20}{9}\,T_F\, n_f \Bigr]
\,,\nn\\
\Gamma_2 &= 4 C_F \Bigl[
C_A^2 \Bigl(\frac{245}{6} -\frac{134 \pi^2}{27} + \frac{11 \pi ^4}{45}
  + \frac{22 \zeta_3}{3}\Bigr)
  + C_A\, T_F\,n_f \Bigl(- \frac{418}{27} + \frac{40 \pi^2}{27}  - \frac{56 \zeta_3}{3} \Bigr)
\nn\\* & \hspace{8ex}
  + C_F\, T_F\,n_f \Bigl(- \frac{55}{3} + 16 \zeta_3 \Bigr) 
  - \frac{16}{27}\,T_F^2\, n_f^2 \Bigr]
\,, \\[2ex]
\gamma_{H\,0} &= -12 C_F
\,,\nn\\
\gamma_{H\,1}
&= C_F \Bigl[
  C_A \Bigl(-\frac{164}{9} +104 \zeta_3\Bigr) 
+ C_F (-6 +8 \pi^2 - 96 \zeta_3) 
+ \beta_0 \Bigl(-\frac{130}{9} -2 \pi^2 \Bigr) \Bigr]
\,,\\[2ex]
\gamma_{J\,0} &= 6 C_F
\,,\nn\\
\gamma_{J\,1}
&= C_F \Bigl[
  C_A \Bigl(\frac{146}{9} - 80 \zeta_3\Bigr) 
+ C_F (3 - 4 \pi^2 + 48 \zeta_3) 
+ \beta_0 \Bigl(\frac{121}{9} + \frac{2\pi^2}{3} \Bigr)  \Bigr]
\,,\\[2ex]
\gamma_{S_{\rm thr}\,0} &= 0
\,,\nn\\
\gamma_{S_{\rm thr}\,1}
&= C_F \Bigl[
   C_A \Bigl( -\frac{128}{9} + 56 \zeta_3 \Bigr)
 + \beta_0 \Bigl( -\frac{112}{9} + \frac{2\pi^2}{3} \Bigr)
\Bigr]
\,,\\[2ex]
\gamma_{S\,0} &= 0
\,,\nn\\
\gamma_{S\,1}
&= 0
\,,\\[2ex]
\ga_{\CS\,0} &= 0
\,,\nn\\
\ga_{\CS\,1} &= C_F \Bigl[C_A\,\Bigl( \frac{128}{9} - 56 \zeta_3 \Bigr)  + \beta_0 \Bigl(\frac{112}{9} - \frac{2\pi^2}{3} \Bigr) \Bigr]
\,,\\[2ex]
\gamma_{D\,0} &= 6 C_F
\,,\nn\\
\gamma_{D\,1}
&= C_F \Bigl[
  C_A (2-24 \zeta_3)
+ C_F (3 - 4 \pi^2 + 48 \zeta_3) 
+ \beta_0 \Bigl(1 + \frac{4\pi^2}{3} \Bigr)  \Bigr]
\,. \end{align}

\bibliographystyle{JHEP}
\normalbaselines 
\bibliography{bibliography} 

\providecommand{\href}[2]{#2}\begingroup\raggedright\begin{thebibliography}{10}

\bibitem{Georgi:1977mg}
H.~Georgi and H.~Politzer, {\it {Quark Decay Functions and Heavy Hadron
  Production in QCD}},  {\em Nucl. Phys. B} {\bf 136} (1978) 445--460.

\bibitem{Ellis:1978ty}
R.~Ellis, H.~Georgi, M.~Machacek, H.~Politzer, and G.~G. Ross, {\it
  {Perturbation Theory and the Parton Model in QCD}},  {\em Nucl. Phys. B} {\bf
  152} (1979) 285--329.

\bibitem{Curci:1980uw}
G.~Curci, W.~Furmanski, and R.~Petronzio, {\it {Evolution of Parton Densities
  Beyond Leading Order: The Nonsinglet Case}},  {\em Nucl. Phys. B} {\bf 175}
  (1980) 27--92.

\bibitem{Collins:1981uw}
J.~C. Collins and D.~E. Soper, {\it {Parton Distribution and Decay Functions}},
   {\em Nucl. Phys. B} {\bf 194} (1982) 445--492.

\bibitem{Collins:1992kk}
J.~C. Collins, {\it {Fragmentation of transversely polarized quarks probed in
  transverse momentum distributions}},  {\em Nucl. Phys. B} {\bf 396} (1993)
  161--182, [\href{http://arxiv.org/abs/hep-ph/9208213}{{\tt hep-ph/9208213}}].

\bibitem{Mulders:1995dh}
P.~Mulders and R.~Tangerman, {\it {The Complete tree level result up to order
  1/Q for polarized deep inelastic leptoproduction}},  {\em Nucl. Phys. B} {\bf
  461} (1996) 197--237, [\href{http://arxiv.org/abs/hep-ph/9510301}{{\tt
  hep-ph/9510301}}]. [Erratum: Nucl.Phys.B 484, 538--540 (1997)].

\bibitem{Boer:1997nt}
D.~Boer and P.~Mulders, {\it {Time reversal odd distribution functions in
  leptoproduction}},  {\em Phys. Rev. D} {\bf 57} (1998) 5780--5786,
  [\href{http://arxiv.org/abs/hep-ph/9711485}{{\tt hep-ph/9711485}}].

\bibitem{Boer:1997qn}
D.~Boer, R.~Jakob, and P.~Mulders, {\it {Leading asymmetries in two hadron
  production in $e^+ e^-$ annihilation at the Z pole}},  {\em Phys. Lett. B}
  {\bf 424} (1998) 143--151, [\href{http://arxiv.org/abs/hep-ph/9711488}{{\tt
  hep-ph/9711488}}].

\bibitem{Ji:2004wu}
X.-d. Ji, J.-p. Ma, and F.~Yuan, {\it {QCD factorization for semi-inclusive
  deep-inelastic scattering at low transverse momentum}},  {\em Phys. Rev. D}
  {\bf 71} (2005) 034005, [\href{http://arxiv.org/abs/hep-ph/0404183}{{\tt
  hep-ph/0404183}}].

\bibitem{Collins:2011zzd}
J.~Collins, {\em {Foundations of perturbative QCD}}, vol.~32.
\newblock Cambridge University Press, 11, 2013.

\bibitem{Metz:2016swz}
A.~Metz and A.~Vossen, {\it {Parton Fragmentation Functions}},  {\em Prog.
  Part. Nucl. Phys.} {\bf 91} (2016) 136--202,
  [\href{http://arxiv.org/abs/1607.02521}{{\tt arXiv:1607.02521}}].

\bibitem{Anselmino:2020vlp}
M.~Anselmino, A.~Mukherjee, and A.~Vossen, {\it {Transverse spin effects in
  hard semi-inclusive collisions}},  {\em Prog. Part. Nucl. Phys.} {\bf 114}
  (2020) 103806, [\href{http://arxiv.org/abs/2001.05415}{{\tt
  arXiv:2001.05415}}].

\bibitem{Sun:2013hua}
P.~Sun and F.~Yuan, {\it {Transverse momentum dependent evolution: Matching
  semi-inclusive deep inelastic scattering processes to Drell-Yan and W/Z boson
  production}},  {\em Phys. Rev. D} {\bf 88} (2013), no.~11 114012,
  [\href{http://arxiv.org/abs/1308.5003}{{\tt arXiv:1308.5003}}].

\bibitem{Bacchetta:2017gcc}
A.~Bacchetta, F.~Delcarro, C.~Pisano, M.~Radici, and A.~Signori, {\it
  {Extraction of partonic transverse momentum distributions from semi-inclusive
  deep-inelastic scattering, Drell-Yan and Z-boson production}},  {\em JHEP}
  {\bf 06} (2017) 081, [\href{http://arxiv.org/abs/1703.10157}{{\tt
  arXiv:1703.10157}}]. [Erratum: JHEP 06, 051 (2019)].

\bibitem{Scimemi:2019cmh}
I.~Scimemi and A.~Vladimirov, {\it {Non-perturbative structure of
  semi-inclusive deep-inelastic and Drell-Yan scattering at small transverse
  momentum}},  {\em JHEP} {\bf 06} (2020) 137,
  [\href{http://arxiv.org/abs/1912.06532}{{\tt arXiv:1912.06532}}].

\bibitem{Bertone:2019nxa}
V.~Bertone, I.~Scimemi, and A.~Vladimirov, {\it {Extraction of unpolarized
  quark transverse momentum dependent parton distributions from
  Drell-Yan/Z-boson production}},  {\em JHEP} {\bf 06} (2019) 028,
  [\href{http://arxiv.org/abs/1902.08474}{{\tt arXiv:1902.08474}}].

\bibitem{Bacchetta:2019sam}
A.~Bacchetta, V.~Bertone, C.~Bissolotti, G.~Bozzi, F.~Delcarro, F.~Piacenza,
  and M.~Radici, {\it {Transverse-momentum-dependent parton distributions up to
  N$^{3}$LL from Drell-Yan data}},  {\em JHEP} {\bf 07} (2020) 117,
  [\href{http://arxiv.org/abs/1912.07550}{{\tt arXiv:1912.07550}}].

\bibitem{Echevarria:2014xaa}
M.~G. Echevarria, A.~Idilbi, Z.-B. Kang, and I.~Vitev, {\it {QCD Evolution of
  the Sivers Asymmetry}},  {\em Phys. Rev. D} {\bf 89} (2014) 074013,
  [\href{http://arxiv.org/abs/1401.5078}{{\tt arXiv:1401.5078}}].

\bibitem{Anselmino:2015sxa}
M.~Anselmino, M.~Boglione, U.~D'Alesio, J.~Gonzalez~Hernandez, S.~Melis,
  F.~Murgia, and A.~Prokudin, {\it {Collins functions for pions from SIDIS and
  new $e^+e^-$ data: a first glance at their transverse momentum dependence}},
  {\em Phys. Rev. D} {\bf 92} (2015), no.~11 114023,
  [\href{http://arxiv.org/abs/1510.05389}{{\tt arXiv:1510.05389}}].

\bibitem{Kang:2015msa}
Z.-B. Kang, A.~Prokudin, P.~Sun, and F.~Yuan, {\it {Extraction of Quark
  Transversity Distribution and Collins Fragmentation Functions with QCD
  Evolution}},  {\em Phys. Rev. D} {\bf 93} (2016), no.~1 014009,
  [\href{http://arxiv.org/abs/1505.05589}{{\tt arXiv:1505.05589}}].

\bibitem{Callos:2020qtu}
D.~Callos, Z.-B. Kang, and J.~Terry, {\it {Extracting the Transverse Momentum
  Dependent Polarizing Fragmentation Functions}},
  \href{http://arxiv.org/abs/2003.04828}{{\tt arXiv:2003.04828}}.

\bibitem{Cammarota:2020qcw}
{\bf Jefferson Lab Angular Momentum} Collaboration, J.~Cammarota, L.~Gamberg,
  Z.-B. Kang, J.~A. Miller, D.~Pitonyak, A.~Prokudin, T.~C. Rogers, and
  N.~Sato, {\it {Origin of single transverse-spin asymmetries in high-energy
  collisions}},  {\em Phys. Rev. D} {\bf 102} (2020), no.~5 054002,
  [\href{http://arxiv.org/abs/2002.08384}{{\tt arXiv:2002.08384}}].

\bibitem{Accardi:2012qut}
A.~Accardi et~al., {\it {Electron Ion Collider: The Next QCD Frontier}:
  {Understanding the glue that binds us all}},  {\em Eur. Phys. J. A} {\bf 52}
  (2016), no.~9 268, [\href{http://arxiv.org/abs/1212.1701}{{\tt
  arXiv:1212.1701}}].

\bibitem{Gutierrez-Reyes:2018qez}
D.~Gutierrez-Reyes, I.~Scimemi, W.~J. Waalewijn, and L.~Zoppi, {\it {Transverse
  momentum dependent distributions with jets}},  {\em Phys. Rev. Lett.} {\bf
  121} (2018), no.~16 162001, [\href{http://arxiv.org/abs/1807.07573}{{\tt
  arXiv:1807.07573}}].

\bibitem{Liu:2018trl}
X.~Liu, F.~Ringer, W.~Vogelsang, and F.~Yuan, {\it {Lepton-jet Correlations in
  Deep Inelastic Scattering at the Electron-Ion Collider}},  {\em Phys. Rev.
  Lett.} {\bf 122} (2019), no.~19 192003,
  [\href{http://arxiv.org/abs/1812.08077}{{\tt arXiv:1812.08077}}].

\bibitem{Gutierrez-Reyes:2019vbx}
D.~Gutierrez-Reyes, I.~Scimemi, W.~J. Waalewijn, and L.~Zoppi, {\it {Transverse
  momentum dependent distributions in $e^+e^-$ and semi-inclusive
  deep-inelastic scattering using jets}},  {\em JHEP} {\bf 10} (2019) 031,
  [\href{http://arxiv.org/abs/1904.04259}{{\tt arXiv:1904.04259}}].

\bibitem{Gutierrez-Reyes:2019msa}
D.~Gutierrez-Reyes, Y.~Makris, V.~Vaidya, I.~Scimemi, and L.~Zoppi, {\it
  {Probing Transverse-Momentum Distributions With Groomed Jets}},  {\em JHEP}
  {\bf 08} (2019) 161, [\href{http://arxiv.org/abs/1907.05896}{{\tt
  arXiv:1907.05896}}].

\bibitem{Arratia:2020nxw}
M.~Arratia, Z.-B. Kang, A.~Prokudin, and F.~Ringer, {\it {Jet-based
  measurements of Sivers and Collins asymmetries at the future Electron-Ion
  Collider}},  \href{http://arxiv.org/abs/2007.07281}{{\tt arXiv:2007.07281}}.

\bibitem{Seidl:2019jei}
{\bf Belle} Collaboration, R.~Seidl et~al., {\it {Transverse momentum dependent
  production cross sections of charged pions, kaons and protons produced in
  inclusive $e^+e^-$ annihilation at $\sqrt{s}=$ 10.58 GeV}},  {\em Phys. Rev.
  D} {\bf 99} (2019), no.~11 112006,
  [\href{http://arxiv.org/abs/1902.01552}{{\tt arXiv:1902.01552}}].

\bibitem{Farhi:1977sg}
E.~Farhi, {\it {A QCD Test for Jets}},  {\em Phys. Rev. Lett.} {\bf 39} (1977)
  1587--1588.

\bibitem{Landry:1999an}
F.~Landry, R.~Brock, G.~Ladinsky, and C.~Yuan, {\it {New fits for the
  nonperturbative parameters in the CSS resummation formalism}},  {\em Phys.
  Rev. D} {\bf 63} (2001) 013004,
  [\href{http://arxiv.org/abs/hep-ph/9905391}{{\tt hep-ph/9905391}}].

\bibitem{Landry:2002ix}
F.~Landry, R.~Brock, P.~M. Nadolsky, and C.~Yuan, {\it {Tevatron Run-1 $Z$
  boson data and Collins-Soper-Sterman resummation formalism}},  {\em Phys.
  Rev. D} {\bf 67} (2003) 073016,
  [\href{http://arxiv.org/abs/hep-ph/0212159}{{\tt hep-ph/0212159}}].

\bibitem{Konychev:2005iy}
A.~V. Konychev and P.~M. Nadolsky, {\it {Universality of the
  Collins-Soper-Sterman nonperturbative function in gauge boson production}},
  {\em Phys. Lett. B} {\bf 633} (2006) 710--714,
  [\href{http://arxiv.org/abs/hep-ph/0506225}{{\tt hep-ph/0506225}}].

\bibitem{DAlesio:2014mrz}
U.~D'Alesio, M.~G. Echevarria, S.~Melis, and I.~Scimemi, {\it {Non-perturbative
  QCD effects in $q_{T}$ spectra of Drell-Yan and Z-boson production}},  {\em
  JHEP} {\bf 11} (2014) 098, [\href{http://arxiv.org/abs/1407.3311}{{\tt
  arXiv:1407.3311}}].

\bibitem{Scimemi:2017etj}
I.~Scimemi and A.~Vladimirov, {\it {Analysis of vector boson production within
  TMD factorization}},  {\em Eur. Phys. J. C} {\bf 78} (2018), no.~2 89,
  [\href{http://arxiv.org/abs/1706.01473}{{\tt arXiv:1706.01473}}].

\bibitem{Ebert:2018gzl}
M.~A. Ebert, I.~W. Stewart, and Y.~Zhao, {\it {Determining the Nonperturbative
  Collins-Soper Kernel From Lattice QCD}},  {\em Phys. Rev. D} {\bf 99} (2019),
  no.~3 034505, [\href{http://arxiv.org/abs/1811.00026}{{\tt
  arXiv:1811.00026}}].

\bibitem{Shanahan:2020zxr}
P.~Shanahan, M.~Wagman, and Y.~Zhao, {\it {Collins-Soper kernel for TMD
  evolution from lattice QCD}},  {\em Phys. Rev. D} {\bf 102} (2020), no.~1
  014511, [\href{http://arxiv.org/abs/2003.06063}{{\tt arXiv:2003.06063}}].

\bibitem{Jain:2011iu}
A.~Jain, M.~Procura, and W.~J. Waalewijn, {\it {Fully-Unintegrated Parton
  Distribution and Fragmentation Functions at Perturbative $k_T$}},  {\em JHEP}
  {\bf 04} (2012) 132, [\href{http://arxiv.org/abs/1110.0839}{{\tt
  arXiv:1110.0839}}].

\bibitem{Boglione:2017jlh}
M.~Boglione, J.~Gonzalez-Hernandez, and R.~Taghavi, {\it {Transverse parton
  momenta in single inclusive hadron production in ${e^ + }{e^ - }$
  annihilation processes}},  {\em Phys. Lett. B} {\bf 772} (2017) 78--86,
  [\href{http://arxiv.org/abs/1704.08882}{{\tt arXiv:1704.08882}}].

\bibitem{Soleymaninia:2019jqo}
M.~Soleymaninia and H.~Khanpour, {\it {Transverse momentum dependent of charged
  pion, kaon, and proton/antiproton fragmentation functions from $e^+e^-$
  annihilation process}},  {\em Phys. Rev. D} {\bf 100} (2019), no.~9 094033,
  [\href{http://arxiv.org/abs/1907.12294}{{\tt arXiv:1907.12294}}].

\bibitem{Boglione:2020cwn}
M.~Boglione and A.~Simonelli, {\it {Universality-breaking effects in $e^+e^-$
  hadronic production processes}},  \href{http://arxiv.org/abs/2007.13674}{{\tt
  arXiv:2007.13674}}.

\bibitem{Kang:2020yqw}
Z.-B. Kang, D.~Y. Shao, and F.~Zhao, {\it {QCD resummation on single hadron
  transverse momentum distribution with the thrust axis}},
  \href{http://arxiv.org/abs/2007.14425}{{\tt arXiv:2007.14425}}.

\bibitem{Dasgupta:2001sh}
M.~Dasgupta and G.~P. Salam, {\it {Resummation of nonglobal QCD observables}},
  {\em Phys. Lett.} {\bf B512} (2001) 323--330,
  [\href{http://arxiv.org/abs/hep-ph/0104277}{{\tt hep-ph/0104277}}].

\bibitem{Bauer:2000ew}
C.~W. Bauer, S.~Fleming, and M.~E. Luke, {\it {Summing Sudakov logarithms in $B
  \to X_s \gamma$ in effective field theory}},  {\em Phys. Rev.} {\bf D63}
  (2000) 014006, [\href{http://arxiv.org/abs/hep-ph/0005275}{{\tt
  hep-ph/0005275}}].

\bibitem{Bauer:2000yr}
C.~W. Bauer, S.~Fleming, D.~Pirjol, and I.~W. Stewart, {\it {An Effective field
  theory for collinear and soft gluons: Heavy to light decays}},  {\em Phys.
  Rev.} {\bf D63} (2001) 114020,
  [\href{http://arxiv.org/abs/hep-ph/0011336}{{\tt hep-ph/0011336}}].

\bibitem{Bauer:2001yt}
C.~W. Bauer, D.~Pirjol, and I.~W. Stewart, {\it {Soft collinear factorization
  in effective field theory}},  {\em Phys. Rev.} {\bf D65} (2002) 054022,
  [\href{http://arxiv.org/abs/hep-ph/0109045}{{\tt hep-ph/0109045}}].

\bibitem{Bauer:2002nz}
C.~W. Bauer, S.~Fleming, D.~Pirjol, I.~Z. Rothstein, and I.~W. Stewart, {\it
  {Hard scattering factorization from effective field theory}},  {\em Phys.
  Rev.} {\bf D66} (2002) 014017,
  [\href{http://arxiv.org/abs/hep-ph/0202088}{{\tt hep-ph/0202088}}].

\bibitem{Beneke:2002ph}
M.~Beneke, A.~P. Chapovsky, M.~Diehl, and T.~Feldmann, {\it {Soft collinear
  effective theory and heavy to light currents beyond leading power}},  {\em
  Nucl. Phys.} {\bf B643} (2002) 431--476,
  [\href{http://arxiv.org/abs/hep-ph/0206152}{{\tt hep-ph/0206152}}].

\bibitem{Bauer:2011uc}
C.~W. Bauer, F.~J. Tackmann, J.~R. Walsh, and S.~Zuberi, {\it {Factorization
  and Resummation for Dijet Invariant Mass Spectra}},  {\em Phys. Rev. D} {\bf
  85} (2012) 074006, [\href{http://arxiv.org/abs/1106.6047}{{\tt
  arXiv:1106.6047}}].

\bibitem{Procura:2014cba}
M.~Procura, W.~J. Waalewijn, and L.~Zeune, {\it {Resummation of
  Double-Differential Cross Sections and Fully-Unintegrated Parton Distribution
  Functions}},  {\em JHEP} {\bf 02} (2015) 117,
  [\href{http://arxiv.org/abs/1410.6483}{{\tt arXiv:1410.6483}}].

\bibitem{Larkoski:2015zka}
A.~J. Larkoski, I.~Moult, and D.~Neill, {\it {Non-Global Logarithms,
  Factorization, and the Soft Substructure of Jets}},  {\em JHEP} {\bf 09}
  (2015) 143, [\href{http://arxiv.org/abs/1501.04596}{{\tt arXiv:1501.04596}}].

\bibitem{forwardcite}
Y.~Makris, J.~K. Michel, F.~Ringer, and W.~J. Waalewijn, \emph{in preparation}.

\bibitem{Salam:WTAUnpublished}
G.~Salam, ``{$E_t^\infty$ Scheme}.'' Unpublished.

\bibitem{Bertolini:2013iqa}
D.~Bertolini, T.~Chan, and J.~Thaler, {\it {Jet Observables Without Jet
  Algorithms}},  {\em JHEP} {\bf 04} (2014) 013,
  [\href{http://arxiv.org/abs/1310.7584}{{\tt arXiv:1310.7584}}].

\bibitem{Neill:2016vbi}
D.~Neill, I.~Scimemi, and W.~J. Waalewijn, {\it {Jet axes and universal
  transverse-momentum-dependent fragmentation}},  {\em JHEP} {\bf 04} (2017)
  020, [\href{http://arxiv.org/abs/1612.04817}{{\tt arXiv:1612.04817}}].

\bibitem{Neill:2018wtk}
D.~Neill, A.~Papaefstathiou, W.~J. Waalewijn, and L.~Zoppi, {\it {Phenomenology
  with a recoil-free jet axis: TMD fragmentation and the jet shape}},  {\em
  JHEP} {\bf 01} (2019) 067, [\href{http://arxiv.org/abs/1810.12915}{{\tt
  arXiv:1810.12915}}].

\bibitem{Collins:1984kg}
J.~C. Collins, D.~E. Soper, and G.~F. Sterman, {\it {Transverse Momentum
  Distribution in Drell-Yan Pair and W and Z Boson Production}},  {\em Nucl.
  Phys. B} {\bf 250} (1985) 199--224.

\bibitem{Stewart:2014nna}
I.~W. Stewart, F.~J. Tackmann, and W.~J. Waalewijn, {\it {Dissecting Soft
  Radiation with Factorization}},  {\em Phys. Rev. Lett.} {\bf 114} (2015),
  no.~9 092001, [\href{http://arxiv.org/abs/1405.6722}{{\tt arXiv:1405.6722}}].

\bibitem{Lustermans:2019plv}
G.~Lustermans, J.~K. Michel, F.~J. Tackmann, and W.~J. Waalewijn, {\it {Joint
  two-dimensional resummation in $q_{T}$ and $0$-jettiness at NNLL}},  {\em
  JHEP} {\bf 03} (2019) 124, [\href{http://arxiv.org/abs/1901.03331}{{\tt
  arXiv:1901.03331}}].

\bibitem{Banfi:2002hw}
A.~Banfi, G.~Marchesini, and G.~Smye, {\it {Away from jet energy flow}},  {\em
  JHEP} {\bf 08} (2002) 006, [\href{http://arxiv.org/abs/hep-ph/0206076}{{\tt
  hep-ph/0206076}}].

\bibitem{Weigert:2003mm}
H.~Weigert, {\it {Nonglobal jet evolution at finite $N_c$}},  {\em Nucl. Phys.
  B} {\bf 685} (2004) 321--350,
  [\href{http://arxiv.org/abs/hep-ph/0312050}{{\tt hep-ph/0312050}}].

\bibitem{Schwartz:2014wha}
M.~D. Schwartz and H.~X. Zhu, {\it {Nonglobal logarithms at three loops, four
  loops, five loops, and beyond}},  {\em Phys. Rev.} {\bf D90} (2014), no.~6
  065004, [\href{http://arxiv.org/abs/1403.4949}{{\tt arXiv:1403.4949}}].

\bibitem{Hagiwara:2015bia}
Y.~Hagiwara, Y.~Hatta, and T.~Ueda, {\it {Hemisphere jet mass distribution at
  finite $N_c$}},  {\em Phys. Lett. B} {\bf 756} (2016) 254--258,
  [\href{http://arxiv.org/abs/1507.07641}{{\tt arXiv:1507.07641}}].

\bibitem{Caron-Huot:2015bja}
S.~Caron-Huot, {\it {Resummation of non-global logarithms and the BFKL
  equation}},  {\em JHEP} {\bf 03} (2018) 036,
  [\href{http://arxiv.org/abs/1501.03754}{{\tt arXiv:1501.03754}}].

\bibitem{Becher:2016mmh}
T.~Becher, M.~Neubert, L.~Rothen, and D.~Y. Shao, {\it {Factorization and
  Resummation for Jet Processes}},  {\em JHEP} {\bf 11} (2016) 019,
  [\href{http://arxiv.org/abs/1605.02737}{{\tt arXiv:1605.02737}}].

\bibitem{Chiu:2007dg}
J.-y. Chiu, F.~Golf, R.~Kelley, and A.~V. Manohar, {\it {Electroweak
  Corrections in High Energy Processes using Effective Field Theory}},  {\em
  Phys. Rev. D} {\bf 77} (2008) 053004,
  [\href{http://arxiv.org/abs/0712.0396}{{\tt arXiv:0712.0396}}].

\bibitem{Becher:2010tm}
T.~Becher and M.~Neubert, {\it {Drell-Yan Production at Small $q_T$, Transverse
  Parton Distributions and the Collinear Anomaly}},  {\em Eur. Phys. J. C} {\bf
  71} (2011) 1665, [\href{http://arxiv.org/abs/1007.4005}{{\tt
  arXiv:1007.4005}}].

\bibitem{Chiu:2011qc}
J.-y. Chiu, A.~Jain, D.~Neill, and I.~Z. Rothstein, {\it {The Rapidity
  Renormalization Group}},  {\em Phys. Rev. Lett.} {\bf 108} (2012) 151601,
  [\href{http://arxiv.org/abs/1104.0881}{{\tt arXiv:1104.0881}}].

\bibitem{Chiu:2012ir}
J.-Y. Chiu, A.~Jain, D.~Neill, and I.~Z. Rothstein, {\it {A Formalism for the
  Systematic Treatment of Rapidity Logarithms in Quantum Field Theory}},  {\em
  JHEP} {\bf 05} (2012) 084, [\href{http://arxiv.org/abs/1202.0814}{{\tt
  arXiv:1202.0814}}].

\bibitem{Becher:2011dz}
T.~Becher and G.~Bell, {\it {Analytic Regularization in Soft-Collinear
  Effective Theory}},  {\em Phys. Lett.} {\bf B713} (2012) 41--46,
  [\href{http://arxiv.org/abs/1112.3907}{{\tt arXiv:1112.3907}}].

\bibitem{Echevarria:2015usa}
M.~G. Echevarria, I.~Scimemi, and A.~Vladimirov, {\it {Transverse momentum
  dependent fragmentation function at next-to-next-to-leading order}},  {\em
  Phys. Rev. D} {\bf 93} (2016), no.~1 011502,
  [\href{http://arxiv.org/abs/1509.06392}{{\tt arXiv:1509.06392}}]. [Erratum:
  Phys.Rev.D 94, 099904 (2016)].

\bibitem{Li:2016axz}
Y.~Li, D.~Neill, and H.~X. Zhu, {\it {An Exponential Regulator for Rapidity
  Divergences}},  \href{http://arxiv.org/abs/1604.00392}{{\tt
  arXiv:1604.00392}}.

\bibitem{Frixione:1998dw}
S.~Frixione, P.~Nason, and G.~Ridolfi, {\it {Problems in the resummation of
  soft gluon effects in the transverse momentum distributions of massive vector
  bosons in hadronic collisions}},  {\em Nucl. Phys.} {\bf B542} (1999)
  311--328, [\href{http://arxiv.org/abs/hep-ph/9809367}{{\tt hep-ph/9809367}}].

\bibitem{Abbate:2010xh}
R.~Abbate, M.~Fickinger, A.~H. Hoang, V.~Mateu, and I.~W. Stewart, {\it {Thrust
  at N$^3$LL with Power Corrections and a Precision Global Fit for
  $\alpha_s(m_Z)$}},  {\em Phys. Rev. D} {\bf 83} (2011) 074021,
  [\href{http://arxiv.org/abs/1006.3080}{{\tt arXiv:1006.3080}}].

\bibitem{Ligeti:2008ac}
Z.~Ligeti, I.~W. Stewart, and F.~J. Tackmann, {\it {Treating the b quark
  distribution function with reliable uncertainties}},  {\em Phys. Rev.} {\bf
  D78} (2008) 114014, [\href{http://arxiv.org/abs/0807.1926}{{\tt
  arXiv:0807.1926}}].

\bibitem{Manohar:2003vb}
A.~V. Manohar, {\it {Deep inelastic scattering as $x \to 1$ using soft
  collinear effective theory}},  {\em Phys. Rev. D} {\bf 68} (2003) 114019,
  [\href{http://arxiv.org/abs/hep-ph/0309176}{{\tt hep-ph/0309176}}].

\bibitem{Bauer:2003di}
C.~W. Bauer, C.~Lee, A.~V. Manohar, and M.~B. Wise, {\it {Enhanced
  nonperturbative effects in Z decays to hadrons}},  {\em Phys. Rev. D} {\bf
  70} (2004) 034014, [\href{http://arxiv.org/abs/hep-ph/0309278}{{\tt
  hep-ph/0309278}}].

\bibitem{Korchemsky:1987wg}
G.~Korchemsky and A.~Radyushkin, {\it {Renormalization of the Wilson Loops
  Beyond the Leading Order}},  {\em Nucl. Phys. B} {\bf 283} (1987) 342--364.

\bibitem{Bauer:2003pi}
C.~W. Bauer and A.~V. Manohar, {\it {Shape function effects in $B \to X_s
  \gamma$ and $B \to X_u \ell \bar \nu$ decays}},  {\em Phys. Rev. D} {\bf 70}
  (2004) 034024, [\href{http://arxiv.org/abs/hep-ph/0312109}{{\tt
  hep-ph/0312109}}].

\bibitem{Echevarria:2014rua}
M.~G. Echevarria, A.~Idilbi, and I.~Scimemi, {\it {Unified treatment of the QCD
  evolution of all (un-)polarized transverse momentum dependent functions:
  Collins function as a study case}},  {\em Phys. Rev. D} {\bf 90} (2014),
  no.~1 014003, [\href{http://arxiv.org/abs/1402.0869}{{\tt arXiv:1402.0869}}].

\bibitem{Bain:2016rrv}
R.~Bain, Y.~Makris, and T.~Mehen, {\it {Transverse Momentum Dependent
  Fragmenting Jet Functions with Applications to Quarkonium Production}},  {\em
  JHEP} {\bf 11} (2016) 144, [\href{http://arxiv.org/abs/1610.06508}{{\tt
  arXiv:1610.06508}}].

\bibitem{Aybat:2011zv}
S.~Aybat and T.~C. Rogers, {\it {TMD Parton Distribution and Fragmentation
  Functions with QCD Evolution}},  {\em Phys. Rev. D} {\bf 83} (2011) 114042,
  [\href{http://arxiv.org/abs/1101.5057}{{\tt arXiv:1101.5057}}].

\bibitem{Kang:2017glf}
Z.-B. Kang, X.~Liu, F.~Ringer, and H.~Xing, {\it {The transverse momentum
  distribution of hadrons within jets}},  {\em JHEP} {\bf 11} (2017) 068,
  [\href{http://arxiv.org/abs/1705.08443}{{\tt arXiv:1705.08443}}].

\bibitem{Chien:2020hzh}
Y.-T. Chien, R.~Rahn, S.~Schrijnder~van Velzen, D.~Y. Shao, W.~J. Waalewijn,
  and B.~Wu, {\it {Azimuthal angle for boson-jet production in the back-to-back
  limit}},  \href{http://arxiv.org/abs/2005.12279}{{\tt arXiv:2005.12279}}.

\bibitem{Kang:2012zr}
Z.-B. Kang, S.~Mantry, and J.-W. Qiu, {\it {N-Jettiness as a Probe of Nuclear
  Dynamics}},  {\em Phys. Rev. D} {\bf 86} (2012) 114011,
  [\href{http://arxiv.org/abs/1204.5469}{{\tt arXiv:1204.5469}}].

\bibitem{Kang:2013nha}
D.~Kang, C.~Lee, and I.~W. Stewart, {\it {Using 1-Jettiness to Measure 2 Jets
  in DIS 3 Ways}},  {\em Phys. Rev. D} {\bf 88} (2013) 054004,
  [\href{http://arxiv.org/abs/1303.6952}{{\tt arXiv:1303.6952}}].

\bibitem{Antonelli:1999kx}
V.~Antonelli, M.~Dasgupta, and G.~P. Salam, {\it {Resummation of thrust
  distributions in DIS}},  {\em JHEP} {\bf 02} (2000) 001,
  [\href{http://arxiv.org/abs/hep-ph/9912488}{{\tt hep-ph/9912488}}].

\bibitem{Dasgupta:2002dc}
M.~Dasgupta and G.~P. Salam, {\it {Resummed event shape variables in DIS}},
  {\em JHEP} {\bf 08} (2002) 032,
  [\href{http://arxiv.org/abs/hep-ph/0208073}{{\tt hep-ph/0208073}}].

\bibitem{Dasgupta:2003iq}
M.~Dasgupta and G.~P. Salam, {\it {Event shapes in e+ e- annihilation and deep
  inelastic scattering}},  {\em J. Phys. G} {\bf 30} (2004) R143,
  [\href{http://arxiv.org/abs/hep-ph/0312283}{{\tt hep-ph/0312283}}].

\bibitem{Gaunt:2014xxa}
J.~R. Gaunt and M.~Stahlhofen, {\it {The Fully-Differential Quark Beam Function
  at NNLO}},  {\em JHEP} {\bf 12} (2014) 146,
  [\href{http://arxiv.org/abs/1409.8281}{{\tt arXiv:1409.8281}}].

\end{thebibliography}\endgroup

\end{document}